\documentclass[aps,pre,onecolumn,showpacs,superscriptaddress,groupedaddress,notitlepage,nofootinbib]{revtex4-1}
\usepackage[english]{babel}
\usepackage{amsmath,amssymb}
\usepackage{graphicx}
\usepackage{dcolumn}
\usepackage{bm}
\usepackage{amsmath}
\usepackage{xcolor}
\usepackage[applemac]{inputenc}
\usepackage{mathtools}
\usepackage{accents}
\usepackage[labelformat=empty,caption=false]{subfig}
\usepackage{keyval}
\usepackage{bm}
\usepackage{enumerate}
\usepackage{enumitem}
\usepackage{stackrel}
\newlength{\dhatheight}

\addto{\captionsenglish}{}

\newcommand{\dermixpar}[3]{\frac{\partial^2 #1}{\partial #2 \, \partial #3}}
\newcommand{\twoML}[3]{E_{#1,#2}\!\left(#3\right)}
\newcommand{\ML}[2]{E_{#1}(#2)}
\newcommand*{\diff}[1]{\mathop{}\!\mathrm{d} #1}

\newcommand{\derpar}[2]{\frac{\partial #1}{\partial #2}}
\newcommand{\dersecpar}[2]{\frac{\partial^2 #1}{\partial #2^2}}

\newcommand{\Average}[1]{\left\langle #1 \right\rangle}

\DeclareMathOperator{\sign}{sign}

\renewcommand{\vec}{\bm}
\newcommand{\mathsym}[1]{{}}
\newcommand{\unicode}[1]{{}}
\newcommand{\Andrea}[1]{#1}

\newcommand{\be}{\begin{eqnarray}}
\newcommand{\ee}{\end{eqnarray}}
\newcommand{\D}{\mathrm{d}}

\usepackage[pagewise]{lineno}

\begin{document}

\title[]{Feynman-Kac equation for anomalous processes with space- and time-dependent forces}
\author{Andrea Cairoli}
\altaffiliation[Present address: ]{Department of Bioengineering, Imperial College London, South Kensington Campus, SW7 2AZ, UK}
\author{Adrian Baule}
\email{Corresponding author: a.baule@qmul.ac.uk}
\affiliation{School of Mathematical Sciences, Queen Mary, University of London, Mile End Road, E1 4NS, UK}
\date{\today}
\begin{abstract}
\noindent
Functionals of a stochastic process $Y(t)$ model many physical time-extensive observables, for instance particle positions, local and occupation times or accumulated mechanical work. When $Y(t)$ is a normal diffusive process, their statistics are obtained as the solution of the celebrated Feynman-Kac equation.
This equation provides the crucial link between the expected values of diffusion processes and the solutions of deterministic second-order partial differential equations.
When $Y(t)$ is non-Brownian, e.g., an anomalous diffusive process, generalizations of the Feynman-Kac equation that incorporate power-law or more general waiting time distributions of the underlying random walk have recently been derived. A general representation of such waiting times is provided in terms of a L\'evy process whose Laplace exponent is directly related to the memory kernel appearing in the generalized Feynman-Kac equation. 
The corresponding anomalous processes have been shown to capture nonlinear mean square displacements exhibiting crossovers between different scaling regimes, which have been observed in numerous experiments on biological systems like migrating cells or diffusing macromolecules in intracellular environments. 
However, the case where both space- and time-dependent forces drive the dynamics of the generalized anomalous process has not been solved yet. 
Here, we present the missing derivation of the Feynman-Kac equation in such general case by using the subordination technique. Furthermore, we discuss its extension to functionals explicitly depending on time, which are of particular relevance for the stochastic thermodynamics of anomalous diffusive systems. Exact results on the work fluctuations of a simple non-equilibrium model are obtained. An additional aim of this paper is to provide a pedagogical introduction to L\'evy processes, semimartingales and their associated stochastic calculus, which underlie the mathematical formulation of anomalous diffusion as a subordinated process. 
\end{abstract}


\maketitle

\section{\label{Sec:1}Introduction} 

In experimental applications one typically measures physical observables $W$, whose time evolution is determined by the underlying dynamics of the system, \Andrea{which is} described by some stochastic process $Y$. Such time-extensive quantities are naturally defined as functionals of the process $Y$ in the form:
\begin{equation}
W(t)=\int_0^t U(Y(r),r)\diff{r},
\label{eq:Wfunc}
\end{equation}
where $U(x,t)$ is some prescribed arbitrary function. 
If $Y$ is a normal diffusive process, these functionals have been employed to model many different physical phenomena by choosing the function $U$ suitably either with an explicit time dependence or without it. 
For instance, in the linear case $U(x,t)=x$, with $Y$ interpreted as a particle's velocity, $W$ represents its position and the $Y$--$W$ representation simply describes the stochastic evolution of the system in \Andrea{the} phase space \cite{risken1989fokker}. If instead we choose $U(x,t)=\delta(x)$ and $U(x,t)=\Theta(x)$, $W$ stands for the local and occupation time respectively \cite{darling1957occupation,agmon1984residence,newman1998diffusive,berezhkovskii1998residence,dhar1999residence,godreche2001statistics,majumdar2002local,blanco2003invariance,mazzolo2004properties,barkai2006residence,grebenkov2007residence,dumonteil2016residence}. Other relevant choices are $U(x,t)=x^2$ with Eq.~\eqref{eq:Wfunc} interpreted as a one-dimensional spatial integral, in which case $W$ is interpreted as the variance of a fluctuating interface \cite{majumdar2005brownian}, and $U(x,t)=e^{-\beta\,x}$ with $\beta$ a real positive parameter, which describes the dynamics of integrated stock prices of the Black-Scholes type \cite{yor2012exponential}. 
Another important class of functionals has been introduced in the context of the stochastic thermodynamics of driven small scale systems. Specifically, one is interested in the statistical properties of the accumulated mechanical work done by the system when a non-equilibrium driving is imposed by a time-dependence in a potential $V(x,q(t))$ via some prescribed time-dependent protocol $q(t)$. 
In such a scenario, assuming $Y$ to be the position coordinate, the mechanical work is defined by Eq.~\eqref{eq:Wfunc} with $U(x,t)=\partial_q V(x,q(t))\dot{q}(t)$ \cite{jarzynski1997nonequilibrium,sekimoto1997kinetic,sekimoto1998langevin,sekimoto2010stochastic,seifert2012stochastic}.

In order to obtain the probability density function (PDF) of $W$ one usually considers quantities of the form:
\be
\label{ftilde}
\widehat{P}(p,y,t)=\left<e^{ipW(t)}\delta(y-Y(t))\right>
\ee
for a given initial condition $Y(0)=y_0$. We note that $\widehat{P}$ is the Fourier-transform of the joint PDF of the processes $W$ and $Y$, such that, if one could compute it, the marginal PDF of $W$ would be obtained straightforwardly by making its Fourier inverse transform and subsequently integrating it over all $y$. Equivalently to the direct evaluation of the expected value in Eq.~\eqref{ftilde}, $\widehat{P}$ can be obtained by solving the Feynman-Kac (FK) equation \cite{majumdar2005brownian}:
\be
\label{FK}
\derpar{}{t}\widehat{P}(p,y,t)&=i\,p\,U(y,t)\,\widehat{P}(p,y,t) +\mathcal{L}_{\rm FP}(y,t)\widehat{P}(p,y,t) , 
\ee
where we introduce the most general Fokker-Planck operator for a space-time dependent force and diffusion coefficient: $\mathcal{L}_{\rm FP}(y,t)=-\derpar{}{y}F(y,t)+\frac{1}{2}\dersecpar{}{y}\sigma^2(y,t)$. In the linear case $U(x)=x$, Eq.~\eqref{FK} maps directly onto the Klein-Kramers equation for the joint position-velocity PDF of a Brownian particle \cite{risken1989fokker}.
The relevance of the FK Equation~\eqref{FK} is motivated by the fact that it allows the calculation of expected values over stochastic trajectories $Y$ of the type of Eq.~\eqref{ftilde} in terms of solutions of second-order partial differential equations, i.e., of non-random equations, and vice versa. In the conventional setting, i.e., that of Eq.~\eqref{FK}, the dynamics of $Y(t)$ is described by the Langevin equation:
\be
\label{langevin}
\dot{Y}(t)=F(y,t)+\sigma(y,t)\xi(t),
\ee
where $\xi(t)$ is a Gaussian white noise with zero mean $\left<\xi(t)\right>=0$ and covariance $\left<\xi(t)\xi(t')\right>=\delta(t-t')$ and the It{\^o}-convention is assumed for the multiplicative term $\sigma(y,t)\xi(t)$ (Appendix~\ref{Sec:ItotoAlpha}). Note that the FK equation contains as a special case the Fokker-Planck equation to which it reduces by setting $p=0$ in Eqs.~(\ref{ftilde}, \ref{FK}). Thus, Eq.~\eqref{FK} is the key method to derive the \Andrea{full statistics} of a \Andrea{wide} range of phenomena modeled by the diffusive dynamics of Eq.~\eqref{langevin} \cite{risken1989fokker}.

In recent years, an intense effort has been dedicated to derive generalizations of both the FK and Klein-Kramers equation, that extend beyond the normal diffusive regime into the anomalous one. First results were obtained either by substituting $\xi(t)$ with a L{\'e}vy noise, such that $Y$ describes L{\'e}vy flight type dynamics, \Andrea{\cite{peseckis1987statistical,fogedby1994levy,jespersen1999levy,lutz2001fractional,eliazar2003levy}} or by directly introducing temporal memory integral terms manifest in time fractional operators into the ordinary FK and Klein-Kramers equations, thus accounting for the non-Markovian effects often characterizing anomalous diffusive processes on a purely phenomenological level \cite{metzler2000generalized1,metzler2000subdiffusive,metzler2000generalized2,barkai2000fractional,metzler2002superdiffusive,zoia2012discrete,fa2013generalized}. 
These fractional FK equations have been successfully used to model, e.g., the dynamics of migrating epithelial cells \cite{dieterich2008anomalous} and the advection of a fluid particle in turbulence \cite{baule2006investigation}. 
However, the relation between such equations and the underlying stochastic dynamics is often not clear \cite{eule2007langevin,eule2012langevin}. Thus, more systematic approaches have been adopted, which explicitly assume the process $Y$ to represent a continuous time random walk (CTRW) with jump lengths and waiting times drawn from independent distributions \cite{montroll1965random,metzler2000random}. 
Specifically, in \cite{Friedrich2006Anomalous,Friedrich2006Exact}, starting from a random walk description of the CTRW in phase space, a Klein-Kramers equation containing a fractional substantial derivative, which generalizes the ordinary material derivative through the inclusion of explicit retardation effects, was derived. In \cite{turgeman2009fractional} a fractional FK equation with the same fractional derivative was derived within a similar random walk description of CTRWs with power-law distributed waiting times. Extensions of this approach to space- and space-time-dependent forces have also been discussed in \cite{carmi2010distributions,carmi2011fractional}, as well as to inhomogeneous media in \cite{shkilev2012equations,shkilev2016feynman}. 

\Andrea{Even if} these equations are systematic extensions of the conventional FK formula, they do not establish the same correspondence between some anomalous stochastic dynamics and the solutions of fractional partial differential equations as in the conventional picture of Eqs.~(\ref{FK}, \ref{langevin}). Instead of using a Langevin-type equation to describe the dynamics of the underlying CTRW, the time evolution of $\widehat{P}$ is derived directly by means of a generalized master equation. 
Such full correspondence has been established only recently in our work \cite{cairoli2015anomalous} by using a general representation of the CTRW in terms of a random time change (also called subordination) of a normal diffusive process. This approach allows in particular to capture straightforwardly different waiting time distributions of the CTRW by a monotonically increasing L\'evy process in an auxiliary time variable. The characteristic Laplace exponent $\Phi$ of the L\'evy process is \Andrea{naturally} related to the memory kernel appearing in the generalized FK equation.  
Our FK formula has been recently confirmed in \cite{wu2016tempered}, within a random walk approach, for the special case of tempered L{\'e}vy-stable distributed waiting times. 

As shown in \cite{cairoli2015anomalous}, by employing a variable parametric form of $\Phi$, one can fit the resulting anomalous process to mean-square displacement \Andrea{(MSD)} data displaying a nonlinear crossover between, e.g., subdiffusive and normal diffusive scaling regimes.
In addition, the quantitative form of the higher-order correlation functions of both the CTRW and its observables are fully specified for general $\Phi$, such that they can be readily compared with the experimental data to assess the nature of the underlying stochastic process. Evidence of such crossover scaling behavior has been found in a large number of recent experiments of diffusion in biophysical systems ranging from migrating and foraging cells \cite{dieterich2008anomalous,selmeczi2005cell,selmeczi2008cell,campos2010persistent,harris2012generalized} to macromolecules and living organelles, e.g., mitochondria, inside the cytoplasm \Andrea{\cite{caspi2000enhanced,levi2005chromatin,brangwynne2007force,bronstein2009transient,bruno2009transition,senning2010actin,Jeon2011InVivo,jeon2012anomalous,weber2012nonthermal,von2013anomalous,tabei2013intracellular,javer2014persistent}.}
Thus, our framework can be applied to a large variety of different systems exhibiting such anomalous diffusive behavior.

Our main purpose in the present manuscript is to extend the derivation of the generalized FK equation for CTRWs with an arbitrary waiting time distribution \cite{cairoli2015anomalous} to both space- and time- dependent forces and explicit time-dependent functionals. These latter ones in particular, to our knowledge, have not yet been considered in previous works on anomalous diffusion processes and their observables, despite their great importance in the context of the stochastic thermodynamics of small scale systems. 
Even though a FK equation for space-time-dependent forces acting on the CTRW has already been presented in \cite{carmi2011fractional} within a random walk approach for power-law distributed waiting times, its extension to arbitrary \Andrea{distributions} as expressed in the subordination approach of \cite{cairoli2015anomalous} has not been presented so far. 
Thus, we here provide the missing link to a comprehensive understanding of functionals of anomalous processes. Specifically, our proposed equations will allow to model the effect of non-equilibrium work protocols on biophysical systems exhibiting complex anomalous diffusion. An additional purpose is to provide a largely self-contained and pedagogical introduction to L\'evy processes, semimartingales and their stochastic calculus, which is necessary to understand the mathematical framework underlying the description of anomalous processes in terms of subordination.

The remainder of this paper is organized as follows. In Sec.~\ref{Sec:subLangevin} we review the definition of the CTRW model with arbitrarily distributed waiting times and its representation in the diffusive limit as a subordinated stochastic process, whose dynamics is described by coupled Langevin equations. In Sec.~\ref{Sec:mathBasis} we provide the mathematical fundamentals that are necessary to manipulate this representation formally. The stochastic calculus of such subordinated processes is briefly discussed. In Sec.~\ref{Sec:derivation} we use the appropriate form of Ito's lemma to derive generalized FK equations. For pedagogical reasons, we first treat the case of a space-dependent force and time-independent functional as in \cite{cairoli2015anomalous}. We then extend the method to space- and time-dependent forces and time-dependent functionals. In Sec.~\ref{Sec:Applications} we apply our results to study the accumulated mechanical work fluctuations in a simple non-equilibrium model with anomalous dynamics. Finally, in Sec.~\ref{Sec:conclusions} we provide some final remarks on open questions and future work.

\section{\label{Sec:subLangevin} Anomalous processes with general waiting times}

The discussion of random walks with arbitrarily distributed waiting times goes back to the seminal work by Montroll and Weiss \cite{montroll1965random}. In this picture, the random walk is defined as a renewal process, where the walker selects the jump lengths as identically and independently distributed (i.i.d.) random variables (RVs). The waiting times between each jumps are also i.i.d. RVs with a distribution that is possibly correlated with the jump length one. In this paper we generally assume that waiting times and jump lengths are uncorrelated. For a discussion of the correlated case, we refer to \cite{tejedor2010anomalous,magdziarz2012correlated,magdziarz2012langevin,schulz2013correlated,de2013flow,liu2013continuous,magdziarz2013asymptotic}. A natural parametrisation of such a random walk is obtained in terms of the number $n$ of jumps performed. 
If we call $\xi_j$ the amplitude of the jump occurring at the $j$th step and by $\eta_j$ the waiting time between the $(j-1)$th and $j$th jumps, the position $Y$ and the elapsed time $T$ are given by summing all such $n$ RVs: 
\begin{align}
Y&=y_0+\sum_{j=1}^{n}\xi_j , & 
T&=\sum_{j=1}^{n}\eta_j ,
\label{eq:FogAdd}
\end{align}
where $y_0$ denotes the initial position. 
Rather than a parametrisation in terms of the discrete variable $n$, it is usually preferable to describe the position coordinate in terms of a continuous time variable $t$. In Eq.~\eqref{eq:FogAdd} we see that $T$ and $n$ are complementary variables, i.e., either one considers $n$ to be a fixed (integer) number, in which case $T$ is a RV, or one considers directly $n$ to be a RV that gives the number of jumps in a time interval $[0,t]$, where $t$ is the elapsed physical time. If one adopts the latter viewpoint, $n$ becomes the stochastic process $N(t)$, \Andrea{which is defined formally as $N(t)=\max{\{n \geq 0: T(n)\leq t\}}$}, and the position variable can be written as below:
\be
\label{yt}
Y(t)=y_0+\sum_{j=1}^{N(t)}\xi_j . 
\label{eq:Ysum}
\ee
Now the waiting time statistics are contained in $N(t)$. 
Consequently, there are two main methods to obtain the statistics of $Y(t)$ from this Montroll-Weiss random walk picture: (i) One can formulate a generalized master equation for the PDF $f_Y(y,t)\!=\!\langle \delta(y-Y(t))\rangle$ directly from Eq.~\eqref{eq:Ysum}. The master equation is then further approximated on a diffusive time and spatial scale leading to Fokker-Planck equations with fractional time derivatives, which describe the time evolution of $f_Y(y,t)$ \cite{metzler1998fractional,metzler1998anomalous,metzler1999deriving,metzler1999anomalousPRL,metzler1999transport,metzler2000random,barkai2000continuous}. (ii) The diffusive limit can already be considered on the level of Eq.~\eqref{eq:FogAdd}, thus leading to a coupled set of Langevin equations describing the stochastic process $Y(t)$ \cite{fogedby1994langevin,baule2005joint,weron2008modeling}. The resulting Fokker-Planck equation for $f_Y(y,t)$ is equivalent to that obtained by using approach (i) \cite{baule2005joint,magdziarz2007fractional,magdziarz2008equivalence,henry2010fractional}. 

\begin{figure}[!thb]
\centering
\includegraphics[scale=0.14]{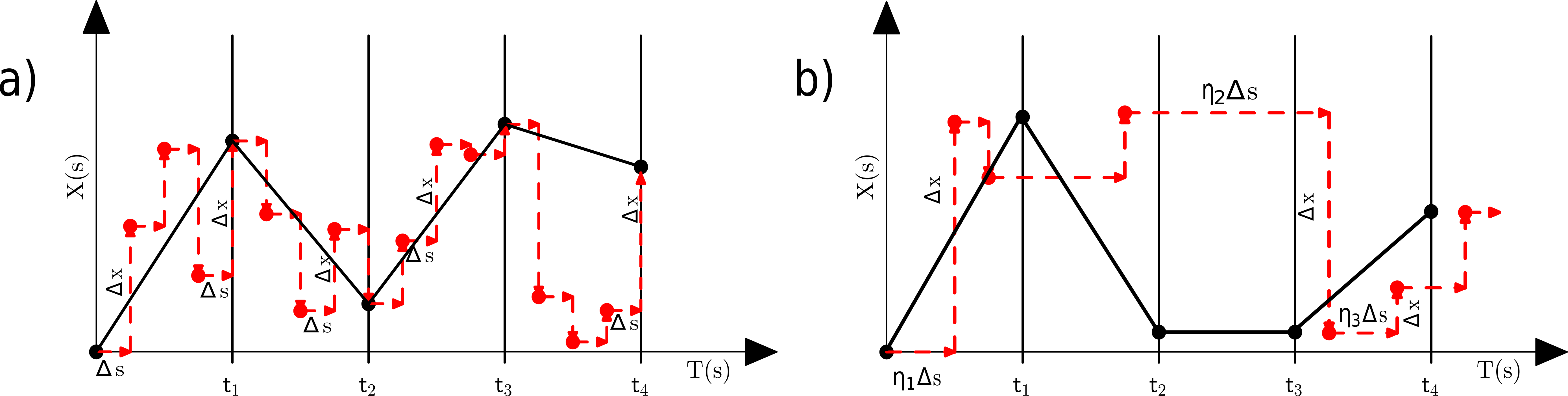}
\caption{Illustration of stochastic processes defined by a subordination as in Eqs.~(\ref{eq:add2},\ref{eq:hitting}). 
We consider (i) a discretisation of the physical time $t$ (black solid lines) of step length $\Delta t$ and (ii) a discretisation of finite step length $\Delta s \ll \Delta t$ of $s$ (not shown). 
We denote with $\Delta x$, $\Delta T$ the increments of the processes $X$ and $T$ corresponding to an increment $\Delta s$ (red dots), which are given by Eqs.~(\ref{eq:FogsubLEs}).
The resulting process $Y(t)=X(S(t))$, with $S$ defined by Eq.~\eqref{eq:hitting}, is plotted in black solid lines. The process $X(s)$ is plotted in dashed red lines and considered as an ordinary random walk. \Andrea{$\xi$(s) is thus a white Gaussian noise.} 
(a) Normal Diffusion. In this case $T$ is a deterministic drift and $\Delta T=\Delta s$. Therefore, $s$ coincides with the physical time $t$ and its discretisation is a thinner time partition.
(b) Anomalous Diffusion. In this case, $\Delta T=\eta_j\,\Delta s$, with $\eta_j$ being a RV.
Consequently, $s$ no longer coincides with the physical time, but it provides a parametrisation of the elapsed physical time $T$ via Eq.~\eqref{eq:hitting}. We note the occurrence of trapping events ($Y(t_2)=Y(t_3)$ in panel b), which are due to the variable length of $\Delta T$. }
\label{fig:Chap2F1}
\end{figure}

In the following, we focus on (ii) and provide a pedagogical introduction to the mathematical framework that is needed to describe anomalous diffusive systems within this approach. The key step is to take a continuum limit in the number of steps: $N(t)\to S(t)$ \Andrea{\cite{meerschaert2011fractional}}. In such a continuum limit Eqs.~(\ref{eq:FogAdd}) become: 
\begin{align}
X(s)&=y_0+\int_0^s \xi(s)\diff{s}, &
T(s)&=\int_0^s \eta(s)\diff{s}, 
\label{eq:FogT}
\end{align}
where $s$ is now interpreted as an auxiliary or operational time variable. 
The position coordinate Eq.~(\ref{yt}) becomes:
\begin{align}
Y(t)&=\int_0^{S(t)} \xi(\tau) \diff{\tau}=X(S(t)). 
\label{eq:add2}
\end{align}
\Andrea{Thus, one must} distinguish the two processes $Y$ and $X$, which are parametrised by the physical and the auxiliary time respectively.   
The complementary relationship between them is naturally expressed by
\begin{equation}
S(t)=\inf_{s>0}{\left\{s:T(s)>t\right\}},
\label{eq:hitting}
\end{equation} 
i.e., $S$ is defined as a collection of first passage times. Indeed, this definition ensures that $S$ accounts exactly for the number of steps, such that the total elapsed time, i.e., the sum of the waiting time increments for each of those steps, is equal to $t$. We see that Eq.~\eqref{eq:hitting} defines $S$ formally as the inverse process of $T$. 
Thus, the CTRW $Y(t)$ is naturally defined as $Y(t)=X(S(t))$, i.e., as a time-changed or subordinated process. The mathematical details underlying the subordination concept are addressed in Sec.~\ref{Sec:time-changedP}. An illustrative picture of this representation, compared with the ordinary random walk, i.e., a normal diffusion, is presented in Fig.~\ref{fig:Chap2F1}. We note that under the assumption of uncorrelated jump lengths and waiting times the PDF of $Y(t)$ can be expressed in terms of the integral transform:
\begin{align}
\label{fyfact}
f_Y(y,t)=\left<\delta(y-Y(t))\right>&=\left<\int_0^\infty\D s\,\delta(s-S(t))\delta(y-X(s))\right>\nonumber\\
&=\int_0^\infty\D s\,h(s,t)f_X(y,s),
\end{align}
where $f_X$ is the PDF associated with the process $X(s)$ and $h(s,t)$ that of $S(t)$.

Regarding the waiting time process, a widely studied case is that of $T$ being a one-sided L\'evy-stable process of order $0 < \alpha \leq 1$, which corresponds to a distribution of the waiting times with power-law tails and diverging first moment. In this specific case, $T$ has the characteristic function:
\be
\label{Tlevystable}
\left<e^{-\lambda T(s)}\right>=e^{-s\lambda^\alpha}.
\label{eq:CTRWPhi}
\ee
Thus, the time-change $S$ in Eq.~\eqref{eq:hitting} is an inverse L\'evy-stable subordinator with a PDF defined in Laplace space as $\widetilde{h}(s,\lambda)=\int_0^\infty\D t\,e^{-\lambda t}h(s,t)=\lambda^{\alpha -1}\,e^{-s\,\lambda^{\alpha}}$ \cite{baule2005joint}. When $X$ describes a pure diffusion process with noise strength $\sigma$, the MSD of $Y$ exhibits a power-law scaling of the same exponent $\alpha$, i.e., ${\rm MSD}(t)=\left<[Y(t)-y_0]^2\right>=\sigma\, t^\alpha/\Gamma(1+\alpha)$. 
This particular scaling regime is characteristic of a pure subdiffusive system \cite{metzler2000random}.

However, in realistic situations the MSD does not always exhibit a single power-law scaling. In fact, diffusive systems, whose MSD exhibits possibly multiple crossovers between different scaling regimes, are widely observed \cite{selmeczi2005cell,selmeczi2008cell,dieterich2008anomalous,harris2012generalized,campos2010persistent,caspi2000enhanced,levi2005chromatin,brangwynne2007force,bronstein2009transient,bruno2009transition,senning2010actin,jeon2012anomalous,weber2012nonthermal,von2013anomalous,tabei2013intracellular,javer2014persistent}. The generalization of the power-law case to account for such more general MSDs is obtained mathematically by choosing $T$ to be a general one-sided \Andrea{strictly} increasing {\it L\'evy process}. Indeed, such a process satisfies the minimal assumptions needed to assure independent and stationary waiting times and causality of $T$. Thus, we specify $T$ by means of its characteristic function, which is given by \cite{cont1975financial,applebaum2009levy}:
\be
\label{Tlevy}
\left<e^{-\lambda T(s)}\right>=e^{-s\Phi(\lambda)},
\ee
with the Laplace exponent $\Phi$ characterizing the jump structure of the waiting times. Different functional forms of $\Phi$ correspond to different distribution laws of the waiting times and of the renewal process $T$. By choosing $\Phi$ suitably, several different waiting time statistics can be captured, i.e., the anomalous process $Y(t)$ can be modeled according to the observed experimental dynamics. If we choose a power law $\Phi(\lambda)=\lambda^{\alpha}$, we recover Eq.~\eqref{eq:CTRWPhi}, i.e., the CTRW case. If instead $\Phi(\lambda)=\lambda$, $T$ is a deterministic drift, $T=s$, and $Y(t)$ reduces to a normal diffusion (Brownian limit) with exponentially distributed waiting times \cite{metzler2000random}. Details on the mathematical properties of $\Phi$ are discussed in Sec.~\ref{Sec:subordinators}.

For $X$ given as a normal diffusion, the MSD of the anomalous process $Y$ with general waiting times can be computed straightforwardly by employing Eq.~\eqref{fyfact}. Indeed, the inverse of the process $T$ has the PDF in Laplace space: $\widetilde{h}(s,\lambda)=[\Phi(\lambda)/\lambda]\,e^{-s\,\Phi(\lambda)}$ \cite{magdziarz2009langevin}. As the first and second moment of $X$ are given by $\langle X(s) \rangle=y_0$ and $\langle [X(s)]^2 \rangle=y^2_0 + \sigma s$ respectively, those of the time-changed process $Y$ read as
$\langle \widetilde{Y}(\lambda)\rangle=\int_0^{\infty} \widetilde{h}(s,\lambda)\,\langle X(s) \rangle=y_0/\lambda$ 
and
$\langle \widetilde{Y}^2(\lambda) \rangle=\int_0^{\infty} \widetilde{h}(s,\lambda)\,\langle X^2(s) \rangle=[y_0^2 + \sigma/\Phi(\lambda)]/\lambda$. 
Putting these results together, we obtain the MSD in Laplace space:  
\begin{align}
\widetilde{\rm MSD}(\lambda)&=\frac{\sigma}{\lambda\,\Phi(\lambda)} .
\label{eq:MSDfreDGPhi}
\end{align}
We note that for $\Phi(\lambda)=\lambda^{\alpha}$ we recover the single power-law scaling previously discussed. For different choices of $\Phi$ Eq.~\eqref{eq:MSDfreDGPhi} is able to capture many different scaling behaviors of ${\rm MSD}(t)$ \cite{cairoli2015anomalous}. For instance, in the case of $T$ given as a tempered L{\'e}vy-stable process, i.e., $\Phi(\lambda)=(\mu + \lambda)^{\alpha} -\mu^{\alpha}$ with $\mu \in \mathbb{R}^+$, the MSD displays a crossover between a power-law (of exponent $\alpha$) and a normal linear scaling. In the case of a sum of two independent stable distributions with exponents $\alpha_1<\alpha_2$, i.e., $\Phi(\lambda)=C_1\,\lambda^{\alpha_1} + C_2\,\lambda^{\alpha_2}$ with $C_1,C_2 \in \mathbb{R}^+$, the crossover is between two subdiffusive regimes with power-law scaling of exponents $\alpha_1/\alpha_2$ respectively for long/small times \Andrea{\cite{sandev2015distributed}}. 

Equations~\eqref{eq:FogT} allow us to express $X$ and $T$ in terms of Langevin equations, as first formulated by Fogedby \cite{fogedby1994langevin}:
\begin{align}
\dot{X}(s)&=\xi(s), & 
\dot{T}(s)&=\eta(s). \label{eq:FogsubLEs}
\end{align}
with initial conditions $X(0)=Y(0)=y_0$ and $T(0)=0$. 
Even though these steps seem somewhat superfluous, there are various advantages by expressing both $X$ and $T$ in this way. In particular, it allows to easily incorporate forces acting on the random walker during the instantaneous jumps. The case where they affect the dynamics of the walker also during the waiting times is discussed in \cite{fedotov2014sub,cairoli2015langevin}. The factorization of the expected values leading to Eq.~(\ref{fyfact}) still holds in the presence of forces that depend only on the position or on the auxiliary time variable, i.e., when Eq.~\eqref{eq:FogsubLEs}(left) is substituted by $\dot{X}(s)=F(X(s),s)+\xi(s)$. 
However, in realistic scenarios the force should vary in physical time rather than in the auxiliary one, which is only a formal construct to simplify the mathematical description \cite{heinsalu2007use}. 
Instead, the correct Langevin equation substituting Eq.~\eqref{eq:FogsubLEs}(left) is given as $\dot{X}(s)=F(X(s),T(s))+\xi(s)$ \cite{heinsalu2007use,magdziarz2008equivalence,weron2008modeling,magdziarz2009stochastic,magdziarz2009langevin,eule2009subordinated,heinsalu2009fractional,henry2010fractional}. Now the factorization in Eq.~(\ref{fyfact}) breaks down. Indeed, the evolution of the position in the auxiliary time depends on the statistics of both the jumps lengths and waiting times. Therefore, introducing space- and (physical) time-dependent forces naturally induces a coupling between the two corresponding processes.

By specifying the properties of $X$ and $T$, we can capture the fluctuation properties of a large variety of physical systems. In the following, we generally assume $\xi(s)$ to represent a white Gaussian noise with $\Average{\xi(s)}=0$ and $\Average{\xi(s)\xi(s')}=\delta(s-s')$, such that $X$ is a normal diffusive process in the auxiliary time $s$. 
For generality, we also consider a multiplicative noise strength $\sigma(x)$, which can capture, e.g., the effects of geometrical confinement (see \cite{lau2007state} and references therein). Thus, we consider the following set of coupled Langevin equations:
\begin{align}
\dot{X}(s)&=F(X(s),T(s))+\sigma(X(s))\xi(s), &
\dot{T}(s)&=\eta(s), \label{eq:subLE}
\end{align}
where the functions $F(x,t)$ and $\sigma(x)$ satisfy standard conditions \cite{revuz1999continuous}, we adopt the It\^o prescription for the multiplicative term (Appendix~\ref{Sec:ItotoAlpha}) \Andrea{and we assume the same initial conditions of Eqs.~\eqref{eq:FogsubLEs}}.  
\Andrea{The stochastic process $Y$ describing the dynamics of the position coordinate is again obtained by subordination, i.e., by Eqs.~(\ref{eq:add2}, \ref{eq:hitting})}.

We note that instead of defining $T$ directly by Eq.~\eqref{Tlevy}, it is convenient to specify the underlying noise process $\eta(s)$ in Eq.~\eqref{eq:subLE}(right) by means of its characteristic functional \cite{van1992stochastic,cont1975financial}:
\begin{equation}
G[k(s)]=\left\langle e^{\,-\int_{0}^{\infty}k(s)\eta(s)\diff{s}}\right\rangle=e^{\,-\int_{0}^{\infty}\Phi(k(s))\diff{s}},
\label{eq:LevyChF}
\end{equation}
which specifies the statistics of the whole noise trajectory. Recalling Eq.~(\ref{eq:FogT}), we recover the characteristic function $\left<e^{-\lambda T(s)}\right>$ from Eq.~(\ref{eq:LevyChF}) by setting $k(s')=\lambda\,\Theta(s-s^{\prime})$. This result, together with Eq.~(\ref{eq:subLE})(right), elucidates the renewal nature of the process $T$.  
Such a process is indeed expressed as a sum over waiting time increments $\Delta t=\int_0^{\Delta s}\eta(\tau)\diff{\tau}$ over a small time step $\Delta s$ of characteristic function $\left<e^{-\lambda \Delta t}\right>=e^{-\Delta s\Phi(\lambda)}$, which can be used to simulate the process $Y(t)$ within a suitable discretisation scheme \cite{kleinhans2007continuous}. Specifying the noise $\eta$ by means of its characteristic functional Eq.~(\ref{eq:LevyChF}) also renders the full multi-point statistics of $T$ easily accessible. In the next section we briefly review some fundamental mathematical notions regarding L\'evy processes, semimartingales and time-changed processes. \Andrea{In particular, we characterize the function $\Phi$, which specifies the waiting time statistics of the underlying random walk}. 

\section{\label{Sec:mathBasis}Mathematical basics: L{\'e}vy processes, subordinators and time-changed processes}

In this section, we provide a comprehensive and accessible review of the theory of L\'evy processes, subordinators and time-changed processes, that are necessary to formulate and work with Langevin equations of anomalous diffusive processes. For more details beyond our present discussion we refer to the monographs \cite{cont1975financial,applebaum2009levy}. 

\subsection{\label{Sec:LevyP} L{\'e}vy processes}

A stochastic process $Y(t)$ for $t\geq 0$ and initial condition $Y(0)=y_0$ is a \textit{L{\'e}vy process} if the conditions below hold: 
\begin{enumerate} 
\item{$Y(0)=y_0=0$ almost surely \Andrea{(a.s.)}, i.e., for each of its different realizations.} 
\item{$Y(t)$ has independent increments, i.e., $\forall\, n \geq 2$ and for each partition $0 \leq t_0 < t_1 < \ldots < t_n\leq t$ the RVs $\{Y(t_j)-Y(t_{j-1})\}_{j=1,\ldots ,n}$ are independent.} 
\item{$Y(t)$ has stationary increments, meaning that for all $0 \leq t_1 < t_2 \leq t$ the RV $Y(t_2)-Y(t_1)$ has the same distribution as $Y(t_2-t_1)$. Note that, if 1 is not satisfied, it would instead depend on $Y(t_2-t_1)-y_0$.}  
\item{The trajectories of $Y(t)$ are c{\`a}dl{\`a}g, i.e., right continuous with left limits.}  
\end{enumerate}  
If one restricts the conditions 2, 4 by assuming Gaussian distributed increments and continuous trajectories respectively, one recovers ordinary Brownian motion. Moreover, as a consequence of (ii), $Y$ is infinitely divisible $\forall\, t \geq 0$. 

The notion of infinitely divisibility characterizes a RV $Y$ that can be expressed as a sum of different i.i.d. RVs $\{X_j\}$. Specifically, we assume $Y$ to have PDF $f_Y$ and corresponding characteristic function: $\phi_Y(k)=\Average{e^{i\,k\,Y}}=\int_{-\infty}^{+\infty} e^{i\,k\,x}\,f_Y(x)\diff{x}$. 
If $\forall\, n \in \mathbb{N}$ there exist i.i.d RVs $X^{(n)}_1,\ldots,X^{(n)}_n$ with PDF $f_X$ and characteristic function $\phi_X$ (uniquely defined), such that the following relation holds (in distribution): 
\begin{equation}
Y = \sum_{j=1}^n X_j^{(n)} ,
\label{eq:InfDiv}
\end{equation}  
then $Y$ is said to be \textit{infinitely divisible}. 
Thus, their characteristic functions are related by the following equation:
\begin{align}
\phi_Y(k)&=\Average{e^{i\,k\,\sum_{j=1}^n X^{(n)}_j}} 
=\prod_{j=1}^n \Average{e^{i\,k\,X^{(n)}_j}\!} 
=\left[\Average{e^{i\,k\,X_1^{(n)}}}\right]^n =[\phi_X(k)]^n .
\label{eq:InfDibCHF}
\end{align}
Note that the factorization of the average is due to the independence of the RVs $X_j^{(n)}$ and that Eq.~\eqref{eq:InfDibCHF} represents a necessary and sufficient condition for $Y$ to be infinitely divisible \cite{applebaum2009levy}, i.e., it can be used as a criterion to asses the infinitely divisibility of a given RV.

Considering now the L{\'e}vy process $Y$ in continuous time, we find that $\forall\, n \in \mathbb{N}$ and $\forall\, t \geq 0$ it can be rewritten as:
\begin{align}
Y(t)&=\sum_{j=1}^n X^{(n)}_j , & X^{(n)}_j&=Y\!\left(j\frac{t}{n}\right)-Y\!\left(\frac{(j-1)\,t}{n}\right) ,
\label{eq:levyDivCond}
\end{align}   
where $\{X^{(n)}_j\}$ are i.i.d. RVs because of 2-3. 
Therefore, such property uniquely relates the characteristic function  of Y to that of a general infinitely divisible RV. 
Indeed, let us introduce the auxiliary function: 
\begin{equation}
\Psi(k,t)=\ln{\Average{\exp{[i\,k\,Y(t)]}}}.
\end{equation}
By suitably adapting Eq.~\eqref{eq:levyDivCond}, we can write for arbitrary $m,n \in \mathbb{N}$: 
\begin{subequations}
\begin{align}
Y(m)&= Y(1) + [Y(2)-Y(1)] + \ldots + [Y(m)-Y(m-1)]=\sum_{j=1}^{m} [Y(j)-Y(j-1)] , \label{eq:levy1}\\
Y\!\left(m\right)&=Y\!\left(\frac{m}{n}\right) 
+\left[Y\!\left(2\frac{m}{n}\right)-Y\!\left(\frac{m}{n}\right)\right]
+ \ldots + \left[Y\!\left(m\right)-Y\!\left((n-1)\frac{m}{n}\right)\right]  
=\sum_{j=1}^{n} \left[Y\!\left(j\frac{m}{n}\right)-Y\!\left((j-1)\frac{m}{n}\right)\right]. \label{eq:levy2}
\end{align}
\end{subequations}
Further exploiting the property 2, Eqs.~(\ref{eq:levy1}, \ref{eq:levy2}) imply that $Y(m) = m\,Y(1)$ and $Y(m) = n\, Y\!\left(\frac{m}{n}\right)$ (in distribution), such that $\Psi(k,m)$ can be computed exactly. 
Specifically, we obtain the two equivalent equations: 
\begin{subequations}
\begin{align}
\Psi(k,m)&=\ln{\Average{\exp{[i\,k\,m\,Y(1)]}}}=m \ln{\Average{\exp{[i\,k\,Y(1)]}}} , \\ 
\Psi(k,m)&=\ln{\Average{\exp{\left[i\,k\,n\,Y\!\left(\frac{m}{n}\right)\right]}}} 
=n\,\ln{\Average{\exp{\left[i\,k\,Y\!\left(\frac{m}{n}\right)\right]}}} .
\end{align}
\end{subequations}
Here, the factorization of the ensemble average is allowed by the independence of the increments. 
Consequently, the equations $\Psi(k,m)=m\,\Psi(k,1)$ and $\Psi(k,m)=n\,\Psi\!\left(k,m/n\right)$ hold simultaneously. If we combine them, we obtain: 
\begin{equation}
\Psi\!\left(k,\frac{m}{n}\right)=\frac{m}{n}\,\Psi(k,1).
\end{equation}
This relation expresses the characteristic function of Y at the finite time $t=m/n$ in terms of its value at time $t=1$. As it is satisfied for every integer $m,n\,$, it also holds for any real positive $t$, i.e., $\Psi\!\left(k,t\right)=t\,\Psi(k,1)$. Moreover, according to Eq.~\eqref{eq:levyDivCond}, $Y(1)$ is an infinitely divisible RV. 
Therefore, a L\'evy process $Y(t)$ has the characteristic function:  
\begin{align}
\phi_Y(k,t)&=\Average{\exp{[i\,k\,Y(t)]}}=\exp{[G(k)\,t]} , 
\label{eq:InfDivCHF}
\end{align} 
where we set $G(k)=\Psi(k,1)$, with $\Psi(k,1)$ being fully specified as the logarithm of the characteristic function of a general infinitely divisible RV. Such quantity, along with the characteristic function of a general L{\'e}vy process according to Eq.~\eqref{eq:InfDivCHF}, is uniquely characterized by means of the L{\'e}vy-Khintchine representation.
Introducing parameters $b \in \mathbb{R}$ and $\sigma \geq 0$, the L{\'e}vy-Khintchine representation states that the characteristic function of any infinitely divisible RV is of the form $\phi(k)=\exp{[G(k)]}$ with the function $G$ defined below \cite{applebaum2009levy}: 
\begin{align}
G(k)&=i\,b\,k - \frac{1}{2}\sigma\,k^2+\int_{\mathbb{R}/\{0\}}[e^{i\,k\,y}-1-i\,k\,y\,\bm{1}_{|y|<1}(y)]\,\Pi(\diff{y}) ,
\label{eq:LKformula}
\end{align}     
with $\bm{1}_A(y)=1$ for $y \in A$ or $\bm{1}_A(y)=0$ otherwise. In Eq.~\eqref{eq:LKformula} $\Pi$ is a so called L{\'e}vy measure, i.e., a probability measure satisfying the following condition:  
\begin{equation}
\int_{\mathbb{R}/\{0\}} \mathrm{Max}{(|y|^2,1)}\,\Pi(\diff{y}) < \infty .
\label{eq:LevyMC}
\end{equation}   
Thus, any L{\'e}vy process $Y$ is uniquely characterized by the triplet $(b,\sigma,\Pi)$, which determines its characteristic function trough Eqs.~(\ref{eq:InfDivCHF}, \ref{eq:LKformula}). Two examples of L{\'e}vy processes are of fundamental importance: 

\begin{itemize}
\item {\it Brownian motion with drift}. In this case, the characteristic function is given by Eq.~\eqref{eq:InfDivCHF} with the specification: 
\begin{align}
G(k)&= i\,b\,k - \frac{\sigma^2}{2}\,k^2 .
\label{eq:BMLevyExp}
\end{align} 
As we can write $G$ as below:  
\begin{align}
G(k)&=\left[\exp{\left(i\,k\,\frac{b}{n}-\frac{\sigma^2}{2\,n}\,k^2\right)}\right]^n=[G_X(k)]^n, 
\label{eq:GChF}
\end{align}     
where we define $G_X(k)=\exp{[i\,k\,b/n-k^2\sigma^2/(2\,n)]}$, we deduce that $G(k)$ is the characteristic function of an infinitely divisible RV. The RVs $Y^{(n)}_j$ in Eq.~\eqref{eq:InfDiv} are Gaussian distributed with mean $b/n$ and variance $\sigma^2/n$.

\item {\it Compound Poisson process}. A Compound Poisson process $Y(t)$ on the interval $[0,t]$ is defined as
\be
Y(t)=\sum_{j=1}^{N(t)}\xi_j,
\ee
where the $\xi_j$ are i.i.d. RVs with law $f_{\xi}$ and $N(t)$ is a Poisson process characterized by an intensity $\lambda$, i.e.,  $p(N(t)=n)=\exp(-\lambda t)[(\lambda t)^n/n!]$. This represents a pure jump process where jumps of length $\xi_j$, drawn as i.i.d. RVs, occur at time points that are spaced by an exponentially distributed waiting time. Within this picture, $\lambda$ represents the average number of jumps per unit time. 
Comparing it with Eq.~\eqref{eq:Ysum}, we note that the Compound Poisson process can be regarded as the simplest renewal process. Its characteristic function can be calculated in a straightforward way by conditioning on the number of jumps:
\begin{align}
\phi(k,t)=\Average{\exp{\left(i\,k\,\sum_{j=1}^{N(t)}\xi_j\right)}}
&=\Average{\Average{\left.\exp{\left(i\,k\,\sum_{j=1}^{n}\xi_j\right)}\right|N(t)=n}}\notag\\
&=\Average{\left[\int_{-\infty}^{+\infty}\D y\,e^{i\,k\,y}\,f_{\xi}(y)\right]^n} \notag\\
&=\exp(-\lambda t)\sum_{n=0}^\infty\left[\int_{-\infty}^{+\infty}\D y\,e^{i\,k\,y}\,f_{\xi}(y)\right]^n\frac{(\lambda t)^n}{n!} \notag\\
&=\exp\left[\lambda\, t\left(\int_{-\infty}^{+\infty}\D y\,e^{i\,k\,y}\,f_{\xi}(y)-1\right)\right].
\end{align}
Therefore, $Y(t)$ is a L\'evy process with   
\begin{align}
G(k)&= \lambda\, \int_{-\infty}^{+\infty} \left(e^{i\,k\,y} -1\right)\,f_{\xi}(y)\diff{y}.
\label{eq:PoissonLevyExp}
\end{align}
Thus, $G(k)$ is again the characteristic function of an infinitely divisible RV, as Eq.~\eqref{eq:InfDiv} is satisfied by choosing the RVs $Y^{(n)}_j$ to have the characteristic function Eq.~\eqref{eq:PoissonLevyExp} with $\lambda\to\lambda/n$.
\end{itemize}

Comparing these two examples with the L{\'e}vy-Khintchine Eq.~\eqref{eq:LKformula} we note that L\'evy processes can be interpreted intuitively as consisting of three different contributions: A deterministic drift, a continuous normal diffusion and a discontinuous jump process. The Compound Poisson process represents the simplest example of the jump contribution, where the L{\'e}vy measure of the jump amplitudes is just a normalizable PDF: $\Pi(\D y)=\lambda\,f_{\xi}(y)\D y$. The specifications given in Eqs.~(\ref{eq:LKformula}, \ref{eq:LevyMC}) extend this case to a wider class of jump processes, which have possibly non-normalizable length distribution or may possess an infinite intensity of small jumps \cite{gardiner1985handbook}. For physical applications, important examples of such more exotic processes are L{\'e}vy-stable and tempered L{\'e}vy-stable processes \Andrea{\cite{samoradnitsky1994stable}}. 

A L{\'e}vy-stable process is a L{\'e}vy process with stable distributed increments, i.e., $G(k)$ in Eq.~\eqref{eq:InfDivCHF} is the characteristic function of a stable RV, which constitutes a special case of infinitely divisible RVs. Let us consider a RV $Y$ and $n$ independent of its copies $\{Y_j\}_{j=1,\ldots ,n}$. If real-valued sequences of parameters $\{c_n\}_{n\in\mathbb{N}}$ and $\{d_n\}_{n\in\mathbb{N}}$ exist, such that the following relation holds in distribution:
\begin{equation}
\sum_{j=1}^n Y_j = c_n\,Y+d_n ,
\label{eq:stableDef}
\end{equation}
then $Y$ is called a \textit{stable} RV. If $d_n=0$, then $Y$ is \textit{strictly stable}. 
From this definition, it is straightforward to see that (i) $Y$ is infinitely divisible [simply set $X_j^{(n)}=(Y_j-d_n/n)/c_n$ in Eq.~\eqref{eq:InfDiv}] and that (ii) the existence of $Y$ represents a generalization of the central limit theorem. 
Indeed, Eq.~\eqref{eq:stableDef} equivalently states that the sequences of partial sums $\{S_n\}_{n\in\mathbb{N}}$ with $S_n=(Y_1+\ldots +Y_n-d_n)/c_n$ converge in distribution to $Y$. With the choice $c_n=\sigma\,\sqrt{n}$ and $d_n=n\,m$, this is the ordinary central limit theorem and $Y$ is Gaussian distributed with mean $m$ and variance $\sigma^2$. For different choices of $c_n$ and $d_n$, we obtain instead a generalized central limit theorem \cite{gnedenko1954limit}.
However, the only possible choice to satisfy Eq.~\eqref{eq:stableDef} is given by $c_n=\sigma\,n^{1/\alpha}$, with $0<\alpha\leq 2$, also called index of stability of the stable distribution \cite{feller1971introduction}.  
As stable distributions are infinitely divisible, their characteristic function is completely determined by  Eq.~\eqref{eq:LKformula}. In particular, we have two possible characteristics: (i) $(b,\sigma^2,0)$ for $\alpha =2$, implying that $Y$ is Gaussian (mean $b$, variance $\sigma^2$) and (ii) $(b,0,\Pi)$ for $\alpha \neq 2$ with $\Pi$ specified by the following formula (for $c_1,c_2 \geq 0$ and $c_1+c_2>0$):
\begin{equation}
\Pi(\diff{y})=\left\{
\begin{array}{ll}
c_1\,y^{-1-\alpha}\,\diff{y} & \qquad y\in[0,\infty) \\
c_2\,|y|^{-1-\alpha}\,\diff{y} & \qquad y\in(-\infty ,0)
\end{array}
\right.
\label{eq:LevyPi}
\end{equation}
By suitably changing coordinates in Eq.~\eqref{eq:LKformula} \cite{sato1999levy}, we obtain the following characterization of $G(k)$:
\begin{subequations}
\begin{align}
G(k)&=i\,\mu\,k -\frac{1}{2}\sigma^2\,k^2 , & \alpha=2 \label{eq:stableChFa}
 \\
G(k)&= i\,\mu\,k -\sigma^{\alpha}\,|k|^{\alpha}\left[1-i\,\beta\,\sign{(k)}\tan{\left(\frac{\pi\,\alpha}{2}\right)}\right] , & \alpha \neq 1,2 \label{eq:stableChFb}
 \\
G(k)&= i\,\mu\,k -\sigma\,|k|\left[1+i\,\beta\,\frac{2}{\pi}\sign{(k)}\log{(|k|)}\right] , & \alpha =1
\label{eq:stableChFc}
\end{align}
\end{subequations} 
for $\mu \in \mathbb{R}$, $\sigma \geq 0$ and $-1 \leq \beta \leq 1$.  
If $X$ is a symmetric stable RV ($\beta=0$), then the function $G$ reduces to: 
\begin{align}
G(k)&=-\rho^{\alpha}\,|k|^{\alpha}, & 0<\alpha\leq 2
\label{eq:SymmStableChF}
\end{align}
with $\rho=\sigma$ for $0<\alpha<2$ and $\rho=\sigma/\sqrt{2}$ for $\alpha=2$. 

Of particular importance for us are one-sided monotonically increasing L\'evy processes, which can be used to implement a random time change. These processes are called subordinators.

\subsection{\label{Sec:subordinators} Subordinators}

We define a \textit{subordinator} a one-dimensional L{\'e}vy process that is \Andrea{a.s.} non-decreasing. 
Thus, if $T(t)$ for $t\geq 0$ is a subordinator, the following properties hold \Andrea{a.s.}: 
(i) $T(t)\geq 0$, $\forall\, t \geq 0$ and 
(ii) $T(t_1)\leq T(t_2)$, $\forall\, t_1 \leq t_2$. 
According to the discussion in Sec.~\ref{Sec:LevyP}, its characteristic function is determined by Eqs.~(\ref{eq:InfDivCHF}, \ref{eq:LKformula}) for a subclass of characteristic triplets $(b,\sigma,\nu)$ that we need to determine.  
We first note that, if $X(t)$ is a Brownian motion of variance $\sigma^2$, we have: $p(X(t)\geq 0)=1/2=p(X(t) \leq 0)$. For a subordinator instead, we require $p(T(t) < 0)=0$ for all times. 
Thus, a subordinator $T$ cannot have any Gaussian component in its L{\'e}vy symbol, i.e., $\sigma =0$ in Eq.~\eqref{eq:LKformula}. 
In addition, the monotonicity of $T$ implies that no jumps of negative amplitudes nor a negative shift are allowed, thus implying the further conditions: $b \geq 0$ and $\Pi(-\infty,0)=0$. 
Taking these requirements into account, the characteristic function of a general subordinator $T$ is given as \cite{bertoin1999subordinators}: 
\be
\Average{e^{-\lambda\,T(t)}}&=&e^{-t\,\Phi(\lambda)}, \label{eq:LapExpA}\\
\Phi(\lambda)&=&b\,\lambda+\int_0^{+\infty}(1-e^{-\lambda\,y})\,\Pi(\diff{y}) , 
\label{eq:LapExpB}
\ee
where one needs to further assume that $\int_0^{+\infty} \mathrm{Max}{(y, 1)}\, \Pi(\diff{y}) < \infty$. 
$\Phi$ is the Laplace exponent of the subordinator. \Andrea{As suggested in Sec.~\ref{Sec:subLangevin}, $\Phi$ determines the characteristic functional Eq.~\eqref{eq:LevyChF} of the noise $\eta$ appearing in Eq.~\eqref{eq:subLE}(right).} 
We remark that only two parameters define its form, i.e., the characteristics of $T$ are determined by the duplet $(b,\Pi)$.    
Using Eq.~\eqref{eq:InfDivCHF} and Jensen's inequality, one can show that $\Phi(\lambda)$ must be a continuous, non negative, non decreasing and concave function. We also remark that $\Phi(0)=0$. 
In general, one can prove that $\Phi$ is a Bernstein function \cite{schilling2012bernstein,meerschaert2015relaxation}. 
Specific examples of subordinators are reviewed in the following: 
\begin{itemize}
\item{ {\it L{\'e}vy stable subordinator}. A subordinator $T$ is L{\'e}vy stable if it has characteristic duplet $(0,\Pi)$ with 
\begin{equation}
\Pi(\diff{y})=\frac{\alpha}{\Gamma(1-\alpha)}y^{-1-\alpha}\diff{y}.
\end{equation} 
If we substitute it inside Eq.~\eqref{eq:LapExpB}, we obtain the following Laplace exponent: 
\begin{align}
\Phi(\lambda)&=\frac{\alpha}{\Gamma(1-\alpha)}\int_0^{\infty}(1-e^{-\lambda\,y})\,y^{-1-\alpha}\diff{y} 
=\frac{\lambda}{\Gamma(1-\alpha)}\int_0^{\infty}e^{-\lambda\,y}\,y^{-\alpha}\diff{y}
=\lambda^{\alpha}.
\label{eq:LapExpStable}
\end{align} 
}

\item{ {\it Tempered L{\'e}vy stable subordinator}. A subordinator $T$ is tempered L{\'e}vy stable if it has characteristic duplet $(0,\Pi)$ with the L{\'e}vy measure \cite{cont1975financial}: 
\begin{equation}
\Pi(\diff{y})=\frac{\alpha}{\Gamma(1-\alpha)}e^{-\mu\,y}y^{-1-\alpha}\diff{y} \qquad \mu>0.
\end{equation} 
If we substitute it inside Eq.~\eqref{eq:LapExpB}, we obtain the following Laplace exponent: 
\begin{align}
\Phi(\lambda)&=\frac{\alpha}{\Gamma(1-\alpha)} \int_0^{\infty}(1-e^{-\lambda\,y})\,e^{-\mu\,y}\,y^{-1-\alpha}\diff{y} \notag\\
&=\frac{\alpha}{\Gamma(1-\alpha)}\left[ -\int_0^{\infty}(1-e^{-\mu\,y})\,y^{-\alpha}\diff{y} + \int_0^{\infty} (1-e^{-(\lambda + \mu)y})\,y^{-\alpha}\diff{y} \right] 
=(\lambda +\mu)^{\alpha} - \mu^{\alpha} ,
\label{eq:LapExpTempStable}
\end{align} 
where we solved the integrals by employing Eq.~\eqref{eq:LapExpStable}. 
The resulting process $T$ interpolates between a L{\'e}vy stable (of order parameter $\alpha$) and an exponential process. This can be shown by applying the Tauberian theorems, which relate the long(small)-$T$ limit of its distribution to the $\lambda \to 0$($\infty$) limit of its Laplace transform Eq.~\eqref{eq:LapExpA}, or equivalently of the function $\Phi$ derived in Eq.~\eqref{eq:LapExpTempStable}. Specifically, for $\lambda \to \infty$ we find: $\Phi(\lambda) \sim \lambda^{\alpha}$, which recovers the case of a L{\'e}vy stable subordinator [see Eq.~\eqref{eq:LapExpStable}]. For $\lambda \to 0$ instead, we obtain: $\Phi(\lambda) \sim \alpha\,\mu^{\alpha -1}\,\lambda$ , such that we can approximate the characteristic function of $T$ as $\langle e^{-\lambda\,T(t)}\rangle \sim 1- \alpha\,\mu^{\alpha -1}\,t\,\lambda \sim (1+\alpha\,\mu^{\alpha -1}\,t\,\lambda)^{-1}$, which is the Laplace transform of an exponential distribution. 
}
\end{itemize} 

\subsection{\label{Sec:time-changedP}Semimartingales and the stochastic calculus of time-changed processes} 

Let us consider a L{\'e}vy process $X(t)$ and a subordinator $T(t)$. 
Thanks to its monotonicity, $T$ can be employed directly as a random parametrisation of time defining a new time-changed process $Y$ trough the relation: $Y(t)=X(T(t))$. Such process is easily shown to still be a L{\'e}vy process \cite{applebaum2009levy}. 
However, as discussed in Sec.~\ref{Sec:subLangevin}, this is not the situation arising for CTRWs, where their representation by coupled Langevin equations in the diffusive limit involves the inverse of the process $T$, i.e., the process $S$ defined in Eq.~\eqref{eq:hitting}. 
The crucial point is that $S$ is generally {\it not a L\'evy process}. Rather, it is part of a more general class of processes called semimartingales, which also contain L\'evy processes as a special case (see below). An important theorem tracing back to the work of Jacod \cite{jacod714calcul} states that semimartingales (and thus L\'evy processes) subordinated by properly defined time-changes, e.g., the process $S$, are again semimartingales. 
Thus, when we study anomalous diffusion at the level of the Langevin representation of Eqs.~(\ref{eq:subLE}), we need to employ the stochastic calculus of semimartingales. Recalling that the trajectories of $S$ are continuous, because $T$ itself is strictly increasing according to Eq.~\eqref{eq:subLE}(right) \cite{revuz1999continuous,kobayashi2011stochastic}, we can focus only on the subclass of continuous semimartingales.

Let us consider a process $M(t)$ and assume that all the information on $M$ up to a chosen time $s$ is known, i.e., we know $M(s)=m_s$. The process $M$ is a \textit{martingale} if the following relation on its conditional average holds \cite{gardiner1985handbook}: 
\begin{equation}
\Average{M(t)|M(s)=m_s}=m_s .
\label{eq:martingale}
\end{equation}
It is instead called a sub-martingale if $\Average{M(t)|M(s)=m_s}\geq m_s$ or a super-martingale if $\Average{M(t)|M(s)=m_s}\leq m_s$. For instance, the Brownian motion $B(t)$ is a martingale, as one can easily verify by direct computation of Eq.~\eqref{eq:martingale}.    
We define a process $Y$ a \textit{semimartingale} if the following decomposition holds: 
\begin{equation}
Y(t)=M(t)+A(t)
\label{eq:semiM}
\end{equation}
where $M(t)$ and $A(t)$ are a martingale and a finite variation process with c{\`a}dl{\`a}g paths respectively. 
We recall that stochastic integration with respect to semimartingales is well defined \cite{kunita1997stochastic}. 
For the sake of our discussion, we will only present their It{\^o} formula. In the specific case of $Y$ being a continuous semimartingale, this is given by:      
\begin{align}
f(Y(t))-f(Y_0)&=\int_0^t f^{\prime}(Y(\tau))\diff{Y(\tau)}+\frac{1}{2}\int_0^t f^{\prime\prime}(Y(\tau))\diff{[Y,Y]_{\tau}}, 
\label{eq:ItoSemiM}
\end{align}
where $[Y,Y]_{t}$ is the quadratic variation of $Y$ (see Appendix~\ref{Sec:QuadVar} for a review).
The extension of Eq.~\eqref{eq:ItoSemiM} to a M-dimensional semimartingale $\vec{Z}$ reads as:    
\begin{align}
f(\vec{Z}(t))-f(\vec{Z}_0)&\!=\!\sum_{i=1}^M \int_0^t f_{i}^{\prime}(\vec{Z}(\tau))\diff{Z^{(i)}(\tau)} +\frac{1}{2}\sum_{i,j=1}^M \int_0^t f_{i,j}^{\prime\prime}(\bm{Z}(\tau))\diff{[Z^{(i)},Z^{(j)}]_{\tau}}, 
\label{eq:ItoSemiMdim}
\end{align}
where $[Z^{(i)},Z^{(j)}]_{t}$ is the joint quadratic variation of $Z^{(i)}$, $Z^{(j)}$, which is defined analogously to the quadratic variation by substituting the squared increment in Eq.~\eqref{eq:quadVar} with the product of the increments of the two processes.   
We note that both $[Y,Y]_{t}$ and $[Z^{(i)},Z^{(j)}]_{t}$ are continuous increasing processes.  
The joint one also has finite variation paths \cite{kunita1997stochastic}. 
For further properties of semimartingales and their theory of stochastic integration we refer to \cite{kunita1997stochastic}.

Recalling that both $T(s)$ and $S(t)$ are monotonically non decreasing, we can deduce that $S$ is a process of finite variation (see Appendix~\ref{Sec:FVP} for a justification of this statement). This property, on the one hand, classify $S$ generally as a semimartingale [according to Eq.~\eqref{eq:semiM}] and, on the other hand, together with the continuity of its paths, enables us to specify its It{\^o} formula for a general differentiable function $f$ as follows (adapted from Eq.~\eqref{eq:ItoFVP} in Appendix~\ref{Sec:FVP}):     
\begin{equation}
f(S(t))-f(0)=\int_0^{t}\derpar{f}{s}(S(\tau))\diff{S(\tau)}.
\label{eq:ItoS}
\end{equation}
In addition, the monotonicity of the paths of $T$ and $S$ also provides the relation \cite{baule2005joint}: 
\begin{equation}
\Theta(s-S(t))=1-\Theta(t-T(s)).
\label{eq:BauleTheta}
\end{equation}
Thus, if we choose $f(S(t))=\Theta(s-S(t))$ in Eq.~\eqref{eq:ItoS} and we use Eq.~\eqref{eq:BauleTheta}, we obtain:
\begin{equation}
\Theta(t-T(s))=\int_0^{t}\delta(s-S(\tau))\diff{S(\tau)}, 
\end{equation}
or equivalently in its corresponding differential form \cite{cairoli2015anomalous}: 
\begin{equation}
\delta(t-T(s))=\delta(s-S(t))\dot{S}(t). \label{eq:deltaRel}
\end{equation}
We note that $\dot{S}(t)=\lim_{\Delta t \to 0}[S(t+\Delta t)-S(t)]/\Delta t$ is a shorthand notation to denote an integration with respect to the time-change. 
With this definition, we can rewrite the coupled Langevin Eqs.~(\ref{eq:subLE}) as a single time-changed stochastic differential equation \cite{kobayashi2011stochastic}. 
To avoid technicalities related to the jumps of $T$, we first neglect the time dependence of the force term, i.e., $F(x,t) \rightarrow F(x)$ in Eq.~\eqref{eq:subLE}(left). The general case will be addressed in Sec.~\ref{SubSec:time}. 
Thus, we can integrate Eq.~\eqref{eq:subLE}(left) directly to obtain ($X(0)=y_0$):  
\begin{equation}
X(s)-y_0=\int_0^s F(X(\tau))\diff{\tau}+\int_0^s \sigma(X(\tau))\diff{B(\tau)}. 
\end{equation}
If we now apply directly the time change, we obtain the integrated equation for $Y$: 
\begin{align}
\label{YSsub}
Y(t)-y_0&=\int_0^{S(t)} F(X(\tau))\diff{\tau}+\int_0^{S(t)} \sigma(X(\tau))\diff{B(\tau)}.   
\end{align}
In order to proceed, we recall the following two key results valid for time-changed semimartingales. Let $Z$ be a continuous semimartingale and $S$ be given by Eq.~\eqref{eq:hitting} for a subordinator $T$. One can prove the following \cite{kobayashi2011stochastic}:   
\begin{subequations}
\begin{align}
\int_{0}^{S(t)} H(s)\diff{Z(s)}&=\int_0^t H(S(\tau)) \diff{Z(S(\tau))} , \label{eq:ChangeOfVar1} \\
[Z(S(t)),Z(S(t))]_{t}&=[Z,Z]_{S(t)} ,
\label{eq:TimeChangeVQ}
\end{align}
\end{subequations} 
where $H$ is any function that can be integrated with respect to $Z$. 
Applying Eq.~\eqref{eq:ChangeOfVar1} to Eq.~\eqref{YSsub} yields: 
\begin{align}
Y(t)-y_0 &=\int_0^{t} F(X(S(\tau))\diff{S(\tau)} 
+\int_0^{t} \sigma(X(S(\tau))) \diff{B(S(\tau))} \notag\\
&=\int_0^{t} F(Y(\tau))\diff{S(\tau)}+\int_0^{t} \sigma(Y(\tau)) \diff{B(S(\tau))} ,
\end{align}
which can finally be written as a Langevin equation by taking its time derivative:
\begin{align}
\dot{Y}(t)&=F(Y(t))\dot{S}(t)+\sigma(Y(t))\xi(S(t))\dot{S}(t).  \label{eq:Yincrement} 
\end{align}
This equation directly expresses the evolution of the increments of $Y$ in terms of those of the time-change $S$. 
The term $\xi(S(t))\dot{S}(t)$ denotes an increment over the time-changed Brownian motion: $\xi(S(t))\dot{S}(t)=\lim_{\Delta t \to 0} [B(S(t+\Delta t))-B(S(t))]/\Delta t$. 
To justify this result, we recall that increments of the Brownian motion can be written in terms of the noise $\xi$ trough the integral relation: $B(t+\Delta t)-B(t)=\int_t^{t+\Delta t} \xi(\tau)\diff{\tau}$, which leads to the relation $\diff{B(t)}=\xi(t)\diff{t}$ between their differentials in the limit $\Delta t \to 0$.  
Analogously, the increment of the time-changed Brownian motion can be related to $\xi$ by the equation: $B(S(t+\Delta t))-B(S(t))=\int_{S(t)}^{S(t+\Delta t)} \xi(\tau)\diff{\tau}$. 
Moreover, recalling that the paths of $S$ are continuous and monotonically increasing, we find that $\Delta S(t)=S(t+\Delta t)-S(t) \to 0$ in the limit $\Delta t \to 0$, i.e., in such limit we obtain: 
$\diff{B(S(t))}=\xi(S(t))\diff{S(t)}=\xi(S(t))\dot{S}(t)\diff{t}$. 

As discussed earlier in this section, $Y$ can be shown to be a semimartingale, as long as the parent process $X$ in Eq.~\eqref{eq:subLE}(left) is a semimartingale \cite{jacod714calcul}. 
In our specific case, $X$ is a Brownian diffusive process, i.e., it satisfies this property. 
Moreover, thanks to the continuity of the stochastic paths of $S$, both the process $Y$ and its general functional $W$, defined as in Eq.~\eqref{eq:Wfunc}, have continuous trajectories \cite{kobayashi2011stochastic}. 
Thus, the It{\^o} formula of $Y$ is given by Eq.~\eqref{eq:ItoSemiM}, where its quadratic variation can be computed by employing Eqs.~(\ref{eq:ChangeOfVar1}, \ref{eq:TimeChangeVQ}) \cite{kobayashi2011stochastic} and recalling that for $X$ normal diffusive we have: $[X,X]_t = \int_0^t \sigma^2(X(s))\diff{s}$. Thus, we can write the following:  
\begin{align}
[Y,Y]_{t}=[X,X]_{S(t)}&=\int_{0}^{S(t)}\sigma^2(X(\tau))\diff{\tau} \notag\\
&=\int_{0}^{t} \sigma^2(X(S(\tau))\diff{S(\tau)} 
=\int_{0}^{t}\sigma^2(Y(\tau))\dot{S}(\tau)\diff{\tau} . 
\label{eq:YquadVarDer}
\end{align}
We note that this same result was also derived in \cite{magdziarz2010path} with a different approach. 
Finally, Eq.~\eqref{eq:YquadVarDer} leads to the following equation for the infinitesimal increment of the quadratic variation of the process $Y$:
\be
\diff{[Y,Y]_t}=\sigma^2(Y(t))\dot{S}(t).
\label{eq:YquadVar}
\ee

\section{\label{Sec:derivation} Derivation of the generalized Feynman-Kac formula}

With these mathematical preliminaries in place, we can derive the generalized FK equation for quantities $\widehat{P}$ of the form Eq.~\eqref{ftilde}, with $W$ given as the general functional Eq.~\eqref{eq:Wfunc}. The underlying stochastic process $Y$ is assumed to be an anomalous process with general waiting times, described by the coupled Langevin Eqs.~(\ref{eq:subLE}), with the noise $\eta$ specified by its characteristic functional Eq.~\eqref{eq:LevyChF}. For pedagogical reasons, we first consider the case of a purely space-dependent force $F(y)$ in Eq.~\eqref{eq:subLE}(left) and a time independent functional $U(x,t)=U(x)$ in Eq.~\eqref{eq:Wfunc}. In this specific case, a brief discussion of this derivation has been presented previously in \cite{cairoli2015anomalous}. Here, we provide the full details of this calculation. We then discuss its extension to space- and time-dependent forces (Sec.~\ref{SubSec:time}) and time-dependent functionals (Sec.~\ref{Sec:WtimeD}). The latter case results in a set of coupled integro-differential evolution equations for $\widehat{P}$.

\subsection{\label{Sec:SpaceDep} Space-dependent forces}

We here consider the case of a purely space-dependent force $F(y)$ in Eq.~\eqref{eq:subLE}(left) and a time independent functional $U(x,t)=U(x)$ in Eq.~\eqref{eq:Wfunc}. We start from the two dimensional joint process $Z(t)=(Y(t),W(t))$. As suggested in Sec.~\ref{Sec:time-changedP}, the process $Z$ is a semi-martingale with continuous paths, as also $Y$ and $W$. 
Thus, its It\^o formula (for a general smooth function $f$) is obtained by adapting Eq.~\eqref{eq:ItoSemiMdim} \cite{kunita1997stochastic} and is given explicitly by:   
\begin{align}
f(Z(t))&=f(Z_0) + \int_0^t \derpar{}{y}f(Z(\tau))\diff{Y(\tau)} + \int_0^t \derpar{}{w}f(Z(\tau))\diff{W(\tau)} \notag\\
& \quad + \frac{1}{2}\int_0^t \dermixpar{f}{y}{w}(Z(\tau))\diff{[Y,W]_{\tau}}  
+ \frac{1}{2}\int_0^t \dersecpar{}{y}f(Z(\tau))\diff{[Y,Y]_{\tau}}+\frac{1}{2}\int_0^t \dersecpar{}{w}f(Z(\tau))\diff{[W,W]_{\tau}}. 
\end{align}
In order to simplify this equation we need the following ingredients: (i) the time-discretised form of Eq.~\eqref{eq:Yincrement} that expresses the increments of $Y$ in terms of the time-change increments $\diff{S(t)}$; (ii) the differential increment of the quadratic variation of $Y$ in Eq.~\eqref{eq:YquadVar}; (iii) the quadratic variation $[W,W]_t$ and covariation $[Y,W]_t$, which are both null as $W$ is a finite variation process (Appendix~\ref{Sec:QuadVar}). Further recalling from Eq.~\eqref{eq:Wfunc} that $\diff{W}\!(t)=U(Y(t))\diff{t}$, we obtain: 
\begin{align}
f(Z(t))&=f(Z_0)+\int_{0}^{t}\derpar{}{w}f(Z(\tau))U(Y(\tau))\diff{\tau} +\int_{0}^{t}\derpar{}{y}f(Z(\tau))F(Y(\tau))\dot{S}(\tau)\diff{\tau} \notag\\
&\quad +\frac{1}{2}\int_{0}^{t}\dersecpar{}{y}f(Z(\tau))\sigma^{2}(Y(\tau))\dot{S}(\tau)\diff{\tau}
+\int_{0}^{t}\derpar{}{y}f(Z(\tau))\sigma(Y(\tau))\xi(S(\tau))\dot{S}(\tau)\diff{\tau}.
\label{eq:ItoSemiM2}
\end{align}
The equation for the double Fourier transform of the joint PDF $\widehat{P}(p,k,t)$ can be derived by evaluating Eq.~\eqref{eq:ItoSemiM2} for $f(Z(t))=e^{i\,k\,Y(t)+i\,p\,W(t)}$. Specifically, we obtain:
\begin{align}
f(Z(t))&=f(Z_0)+i\,p\int_{0}^{t}f(Z(\tau))U(Y(\tau))\diff{\tau} +i\,k\int_{0}^{t}f(Z(\tau))F(Y(\tau))\dot{S}(\tau)\diff{\tau} \notag\\
&\quad -\frac{k^2}{2}\int_{0}^{t}f(Z(\tau))\sigma^{2}(Y(\tau))\dot{S}(\tau)\diff{\tau} +i\,k\int_{0}^{t}f(Z(\tau))\sigma(Y(\tau))\xi(S(\tau))\dot{S}(\tau)\diff{\tau}.
\label{eq:ItoSemiM3}
\end{align}
Finally, we need to take the ensemble average over the realizations of both $\xi$ and $\eta$, the latter determining the realizations of the process $S$. Within the It\^{o} prescription, the last integral in the rhs of Eq.~\eqref{eq:ItoSemiM3} cancels out. This is briefly proven in the following.    
Let us introduce a finite time-discretisation with mesh $\Delta t$ and let $N=t/\Delta t$. We denote: $Z_i=Z(t_i)$, $S_i=S(t_i)$ and $Y_i=Y(t_i)$. 
The stochastic integral can be written as:  
\begin{equation}
\int_{0}^{t}f(Z(\tau))\sigma(Y(\tau))\cdot\xi(S(\tau))\dot{S}(\tau)\diff{\tau} 
=\lim_{N \to \infty \atop \Delta t \to 0} \sum_{i=0}^{N-1} f(Z_i)\sigma(Y_i)[B(S_{i+1})-B(S_i)].
\label{eq:stochIxi}
\end{equation}
Let us take the average over $\xi$ first. For each fixed realization of $S$ we can then write:
$\Average{f(Z_i)\sigma(Y_i)[B(S_{i+1})-B(S_i)]}=\Average{f(Z_i)\sigma(Y_i)}\Average{B(S_{i+1})-B(S_i)}=0$
which is due to (i) the independence of the increments of $B$, that enables us to factorize the average because both $Z_i$, $Y_i$ only depends on its previous increments, and (ii) to the null first moment of $B$. Thus, the averaged Eq.~\eqref{eq:ItoSemiM3} reduces to the following: 
\begin{align}
\Average{f(Z(t))}&=f(Z_0)+i\,p\int_{0}^{t}\Average{f(Z(\tau))U(Y(\tau))}\diff{\tau} +\Average{\int_{0}^{t}f(Z(\tau))\left[ i\,k\,F(Y(\tau))-\frac{k^2}{2}\sigma^{2}(Y(\tau))\right]\dot{S}(\tau)\diff{\tau}} , \label{eq:ItoSemiM4}
\end{align}
where in the second integral the Fourier transform of the FP operator of Eq.~\eqref{eq:subLE}(left): $\mathcal{L}_{\rm FP}(y)=-\derpar{}{y}F(y)+\frac{1}{2}\dersecpar{}{y}\sigma^2(y)$ appears. 
\Andrea{Further recalling that the inverse Fourier transform of $f(Z(t))$ is equal to $e^{i\,p\,W(t)}\delta(y-Y(t))$ and by using the properties of the delta function, we can derive from Eq.~\eqref{eq:ItoSemiM4}} the following equation for $\widehat{P}(p,y,t)$ of Eq.~\eqref{ftilde}:  
\begin{align}
\derpar{}{t}\widehat{P}(p,y,t)&=i\,p\,U(y)\,\widehat{P}(p,y,t) +\mathcal{L}_{\rm FP}(y)\derpar{}{t}\left\langle \int_{0}^{t}e^{i\,p\,W(\tau)}\delta(y-Y(\tau))\dot{S}(\tau)\diff{\tau}\right\rangle. 
\label{eq:prelimFFK}
\end{align}
To close the equation, we need to relate the averaged stochastic integral in Eq.~\eqref{eq:prelimFFK} to $\widehat{P}(p,y,t)$. 
To this aim, we first write $W$ as a subordinated process with the change of variables $\tau=S(r)$, i.e., $T(\tau)=r$, in Eq.~\eqref{eq:Wfunc}:
\begin{align}
W(t)&=A(S(t)), & A(s)&=\int_0^{s} U(X(\tau)) \eta(\tau)\diff{\tau},
\label{eq:subW}
\end{align}
where the noise $\eta(s)$ explicitly appears from Eq.~\eqref{eq:subLE}(right). 
Thus, by employing the property $1=\int_0^{+\infty}\delta(s-S(t))\diff{s}$ and then considering the same discretisation scheme and notation used to derive Eq.~\eqref{eq:stochIxi}, we obtain: 
\begin{align}
\int_{0}^{t}e^{i\,p\,W(\tau)}\delta(y-Y(\tau))\dot{S}(\tau)\diff{\tau}
&=\int_0^t \left[ \int_0^{+\infty} e^{i\,p\,A(s)}\delta(y-X(s))\delta(s-S(\tau))\diff{s}\right] \dot{S}(\tau)\diff{\tau} \notag\\
&=\lim_{N \to \infty \atop \Delta t \to 0} \sum_{i=0}^{N-1} \left[ \int_0^{+\infty} e^{i\,p\,A(s)}\delta(y-X(s))\delta(s-S_i)\diff{s}\right] (S_{i+1}-S_i) \notag\\
&=\lim_{N \to \infty \atop \Delta t \to 0} \sum_{i=0}^{N-1} \left[ \int_0^{+\infty} e^{i\,p\,A(s)}\delta(y-X(s))\delta(t_i-T(s)) \diff{s}\right](t_{i+1}-t_i) \notag\\
&=\int_0^t \left[ \int_0^{+\infty} e^{i\,p\,A(s)}\delta(y-X(s))\delta(\tau -T(s))\diff{s}\right]\diff{\tau},
\label{eq:GFFKstochI}
\end{align}
where (i) the continuity of the paths of $S$ implies that no jump terms appear in the stochastic integral (Appendix \ref{Sec:FVP}) and (ii) we used Eq.~\eqref{eq:deltaRel} to relate the stochastic increments of $S$ to those of $T$. 
If we take the average over the realizations of the two noises $\eta$ and $\xi$ of Eq.~\eqref{eq:GFFKstochI} and then the time derivative of the resulting expression, we obtain:     
\begin{equation}
\derpar{}{t}\left\langle \int_{0}^{t}e^{i\,p\,W(\tau)}\delta(y-Y(\tau))\dot{S}(\tau)\diff{\tau}\right\rangle 
=\int_0^{+\infty} \Average{e^{i\,p\,A(s)}\delta(y-X(s))\delta(t -T(s))}\diff{s} .
\label{eq:Int1}
\end{equation}
Furthermore, the rhs side of Eq.~\eqref{eq:Int1} can be related in Laplace space to the joint PDF $\widehat{P}(p,y,t)$. By using again the representation of $W$ of Eqs.~\eqref{eq:subW} and the property used to derive Eq.~\eqref{eq:GFFKstochI}, $\widehat{P}(p,y,t)$ can be rewritten as follows:     
\begin{align}
\widehat{P}(p,y,t)
&=
\Average{e^{i\,p\,A(S(t))}\delta(y-X(S(t)))} \notag\\
&= 
\int_0^{+\infty} \Average{e^{i\,p\,A(s)}\delta(y-X(s))\delta(s-S(t))}\diff{s} .
\label{eq:A2}
\end{align}
Written in this form, its Laplace transform can be computed straightforwardly. 
Indeed, recalling Eqs.~(\ref{eq:subLE}, \ref{eq:BauleTheta}), we can derive the Laplace transform of $\delta(s-S(t))$ as follows: 
\begin{align}
\int_0^{+\infty}\delta(s-S(t))e^{-\lambda\,t}\diff{t} &=\derpar{}{s}\int_0^{+\infty}\Theta(s-S(t))e^{-\lambda\,t}\diff{t} \notag\\
&=\derpar{}{s}\int_0^{+\infty}\left[1-\Theta(t-T(s))\right] e^{-\lambda\,t}\diff{t}
=\derpar{}{s}\int_0^{T(s)}e^{-\lambda\,t}\diff{t}
=\eta(s)\,e^{-\lambda\,T(s)} ,
\label{eq:Slaplace}
\end{align}
such that the Laplace transform of Eq.~\eqref{eq:A2} is given by:  
\begin{equation}
\skew{3.5}\widehat{\widetilde{P}}(p,y,\lambda)=\int_0^{+\infty} \Average{\Average{\delta(y-X(s))\,\eta(s)\,e^{-\lambda\,T(s)+i\,p\,A(s)}}}\diff{s}. 
\label{eq:LapP}
\end{equation}   
In Eq.~\eqref{eq:LapP}, we explicitly highlighted that the ensemble average is made over the two different noises $\eta$ and $\xi$, whose independence allows us to change arbitrarily the order in which such averages are performed.   
This flexibility can be readily employed to simplify Eq.~\eqref{eq:LapP} by expressing the $\eta(s)$-dependent part of the integrand as a derivative of the characteristic functional $G[k(l)]$ in Eq.~\eqref{eq:LevyChF}. Indeed, by performing the average with respect to $\eta(s)$ first and recalling that $X$ does not depend on it, this implying that the delta function can be taken out of such average, the only quantity needed to be computed is $\Average{\eta(s)\,\exp{[-\lambda\,T(s)+i\,p\,A(s)]}}$. This can be obtained as follows:    
\begin{align}
\Average{\eta(s)\, e^{-\lambda\,T(s)+i\,p\,A(s)}}
&=\Average{\eta(s)\, e^{-\int_0^s \left[\lambda - i\,p\,U(X(r))\right] \eta(r)\diff{r}}} \notag\\
&=-\frac{1}{\lambda - i\,p\,U(X(s))}\derpar{}{s}\Average{e^{-\int_0^s \left[\lambda - i\,p\,U(X(r))\right] \eta(r)\diff{r}}} \notag\\
&=-\frac{1}{\lambda - i\,p\,U(X(s))}\derpar{}{s} e^{-\int_0^s \Phi\left(\lambda - i\,p\,U(X(r))\right)\diff{r}} \notag\\
&=\frac{\Phi\!\left(\lambda - i\,p\,U(X(s))\right)}{\lambda - i\,p\,U(X(s))} \Average{e^{-\int_0^s \left[\lambda - i\,p\,U(X(r))\right] \eta(r)\diff{r}}} ,
\label{eq:LevyCorr}
\end{align}
where we used the characteristic functional Eq.~\eqref{eq:LevyChF} with the test function $k(l)=\left[\lambda-i\,p\,U(X(l))\right]\Theta(s-l)$. 
Substituting Eq.~\eqref{eq:LevyCorr} back into Eq.~\eqref{eq:LapP}, we derive the following relation:
\begin{equation}
\skew{3.5}\widehat{\widetilde{P}}(p,y,\lambda)=\frac{\Phi\!\left(\lambda - i\,p\,U(y)\right)}{\lambda - i\,p\,U(y)} \int_0^{+\infty}\Average{e^{-\lambda T(s) + i\,p\,A(s)}\delta(y-X(s))}\diff{s}, 
\label{eq:LapP2}
\end{equation}
where now the brackets denote again an average over both $\eta$ and $\xi$. 
We note that the Laplace transform of the rhs of Eq.~\eqref{eq:Int1} is equal to the integral of Eq.~\eqref{eq:LapP2}. 
Thus, by expressing it in terms of $\skew{3.5}\widehat{\widetilde{P}}(p,y,\lambda)$, taking its inverse Laplace transform and substituting it back in Eq.~\eqref{eq:prelimFFK}, we derive the generalized FK formula: 
\begin{align}
\derpar{}{t}\widehat{P}(p,y,t)&=i\,p\,U(y)\,\widehat{P}(p,y,t) +\mathcal{L}_{\rm FP}(y)
\left[\derpar{}{t}-i\,p\,U(y)\,\right]\int_0^{t}\,K(t-\tau)\,e^{i\,p\,U(y)(t-\tau)}\,\widehat{P}(p,y,\tau)\diff{\tau}, \label{eq:FFKE} 
\end{align}
where the memory kernel is related to $\Phi$ trough the following relation (in Laplace space): 
\begin{equation}
\widetilde{K}(\lambda)=\Phi(\lambda)^{-1}.
\label{eq:memKernel}
\end{equation}
Eq.~(\ref{eq:FFKE}) highlights that the non-Markovian features of the underlying anomalous process $Y$ result in a temporal memory that is directly related to both the statistics of the waiting times, expressed by the Laplace exponent $\Phi$, and the $y$-coordinate via the function $U$. Consequently, the integral operator expressing the temporal memory does not commute with the Fokker-Planck operator. In the specific case of $\Phi$ as in Eq.~\eqref{eq:LapExpTempStable}, i.e., $T$ is a one-sided tempered L{\'e}vy stable process, Eq.~(\ref{eq:FFKE}) has also recently been confirmed by using a master equation approach \cite{wu2016tempered}.

\subsection{\label{SubSec:time} Space-time dependent forces} 

In the presence of both space- and time-dependent forces and no multiplicative term, Eq.~\eqref{eq:FFKE} with the substitution $\mathcal{L}_{\rm FP}(y)\to\mathcal{L}_{\rm FP}(y,t)=-\derpar{}{y}F(y,t)+\frac{\sigma}{2}\dersecpar{}{y}$ was already proved in the specific case of CTRWs with power-law waiting times starting from a master equation approach in \cite{carmi2011fractional}. 
As discussed in Sec.~\ref{Sec:subLangevin}, the time dependence in the external force is introduced by making $F$ depend explicitly on $T$, i.e., we consider the general dynamics described by the subordinated Langevin Eqs.~(\ref{eq:subLE}). 
There are two main differences with the time independent case. On the one hand, the processes $X$ and $T$ are no longer independent, such that the previous derivation of Eq.~\eqref{eq:LapP2} does not hold any more. Specifically, the delta function in Eq.~\eqref{eq:LapP} needs to be kept inside the average over the realizations of $\eta$.    
On the other hand, while in the time independent case the stochastic paths of $X$, and consequently those of $F$, have continuous paths, in the case of Eq.~\eqref{eq:subLE}(left), due to the explicit dependence of $X$ on the L{\'e}vy process $T$, both its paths and those of $F$ are generally c{\`a}dl{\`a}g, with random jumps occurring in correspondence to those of $T$. 
Nevertheless, thanks to the finite variation of $T$, both $X$ and the time-changed process $Y$ are still semimartingales. 
Thus, we can integrate Eq.~\eqref{eq:subLE}(left) as below:
\begin{equation}
X(s)-y_0=\int_0^s F(X(\tau),T(\tau))\diff{\tau}+\int_0^s \sigma(X(\tau))\diff{B(\tau)} .
\end{equation}   
We remark that the integral over $F$ is done with respect to a process with finite variation and continuous paths, i.e. a deterministic drift (Lebesgue measure), such that the contribution from the random jumps of $T$ is still null. 
As in the time independent case discussed earlier, we can use directly the time-change to write an equation for $Y$:  
\begin{align}
Y(t)-y_0 &= \int_0^{S(t)} F(X(\tau),T(\tau))\diff{\tau} + \int_0^{S(t)} \sigma(X(\tau)) \diff{B(\tau)} 
\notag\\
&=
\int_0^{t} F(X(S(\tau),T(S(\tau)))\diff{S(\tau)}+\int_0^{t} \sigma(X(S(\tau))) \diff{B(S(\tau))} \notag\\
&=\int_0^{t} F(Y(\tau),\tau)\diff{S(\tau)} + \int_0^{t} \sigma(Y(\tau)) \diff{B(S(\tau))},
\end{align}
where we employed again Eq.~\eqref{eq:ChangeOfVar1}. 
After taking its time derivative, we derive the following equation
\cite{kobayashi2011stochastic}:
\begin{align}
\dot{Y}(t)&=F(Y(t),t)\,\dot{S}(t)+\sigma(Y(t))\xi(S(t))\,\dot{S}(t). \label{eq:YincrementTD} 
\end{align}    
This result elucidates that $Y$ has still continuous paths and that both Eqs.~(\ref{eq:ItoSemiM}, \ref{eq:YquadVar}) still hold. 
Similar arguments as in the previous derivation for the time independent case can be made, leading to the same Eq.~\eqref{eq:prelimFFK} with $\mathcal{L}_{\rm FP}(y,t)$ and the same averaged stochastic integral, which needs to be related to the joint PDF. As already highlighted, the proof of Eq.~\eqref{eq:LapP2} needs a more detailed analysis, as both $X$ and $W$ now depend on the realizations of $\eta$. 
Starting from Eq.~\eqref{eq:LapP}, we can first rewrite the explicit $\eta(s)$ dependence as a time derivative of the exponential function as follows: 
\begin{align}
\skew{3.5}\widehat{\widetilde{P}}(p,y,\lambda)&=\int_0^{+\infty} \Average{e^{-\lambda\,T(s)+i\,p\,A(s)}\eta(s)\delta(y-X(s))}\diff{s} \notag\\
&=\int_0^{+\infty} \Average{\delta(y-X(s))\left[\eta(s)\,e^{-\int_0^s \eta(r)\left[\lambda - i\,p\,U(X(r))\right]\diff{r}}\right]}\diff{s} \notag\\
&=-\frac{1}{\lambda-i\,p\,U(y)}\int_0^{+\infty} \Average{\delta(y-X(s))\derpar{}{s}e^{-\int_0^s \eta(r)\left[\lambda - i\,p\,U(X(r))\right]\diff{r}}} \diff{s} , \label{eq:LapPtimeDep}
\end{align}   
where we use again Eq.~\eqref{eq:subW} and we employ the properties of the delta function to factorize the term $[\lambda -i\,p\,U(y)]^{-1}$ out of the integral. 
Differently from the time independent case, the factors inside the ensemble average can no longer be separated, i.e., we cannot compute directly such term by means of Eq.~\eqref{eq:LevyChF}. 
Nevertheless, such expression can still be 
simplified if we look at its discretised form. We consider a partition $\pi=\{0=s_0<s_1<\ldots<s_n=s\}$ of the interval $[0,s]$ with constant mesh $\Delta s$ and $n=s/\Delta s$. We denote: $X(s_i)=X_i$ and $\eta(s_i)=\eta_i$. We recall that $\eta_i$ are RVs with characteristic function specified by $\Phi$, such that $\Delta T_i=\eta_i \, \Delta s$ is the corresponding $T$ increment (according to Eq.~\eqref{eq:subLE}(right)).    
As the delta function in Eq.~\eqref{eq:LapPtimeDep} only imposes a condition on the final point, we can write:
\begin{widetext}
\begin{align}
\Average{\delta(y-X(s))\derpar{}{s}e^{-\int_0^s \eta(r)\left[\lambda - i\,p\,U(X(r))\right]\diff{r}}} 
&=\lim_{n \to \infty \atop \Delta s \to 0}\left.\Average{\frac{e^{-\sum_{j=1}^{n+1}\eta_j\,\left[\lambda -i\,p\,U(X_{j-1})\right]\Delta s}-e^{-\sum_{j=1}^{n}\eta_j\,\left[\lambda -i\,p\,U(X_{j-1})\right]\Delta s}}{\Delta s}}\right|_{X_n=y} \notag\\
&=\lim_{n \to \infty \atop \Delta s \to 0}\left.\Average{e^{-\sum_{j=1}^{n}\eta_j\,\left[\lambda -i\,p\,U(X_{j-1})\right]\Delta s}\left[\frac{e^{-\Delta s\,\eta_{n+1} \left[ \lambda - i\,p\,U(X_n)\right]}-1}{\Delta s}\right]}\right|_{X_n=y} \notag\\
&=\lim_{n \to \infty \atop \Delta s \to 0}\left.\Average{e^{-\sum_{j=1}^{n}\eta_j\,\left[\lambda -i\,p\,U(X_{j-1})\right]\Delta s}}\right|_{X_n=y} \Average{\frac{e^{-\Delta s\,\eta_{n+1} \left[ \lambda - i\,p\,U(y)\right]}-1}{\Delta s}} ,
\label{eq:discretisedLapP2}
\end{align}  
\end{widetext}
where in the first line we discretise the derivative in the operational time $s$ and in the third one we factorized the average over the last increment $\Delta T=\Delta s\,\eta_{n+1}$. This is allowed because (i) $U(X_{n-1})$ only depends on the increments of the process $T$ up to $n$, which are independent on the RV $\eta_{n+1}$, and (ii) the end-point value is conditioned to $y$, i.e. $U(X_{n})=U(y)$, which is no longer a RV. 
The average is then computed with Eq.~\eqref{eq:LevyChF}:
\begin{equation} 
\Average{\frac{e^{-\Delta s\,\eta_{n+1} \left[ \lambda - i\,p\,U(y)\right]}-1}{\Delta s}}=\frac{e^{-\Delta s\,\Phi(\lambda - i\,p\,U(y))}-1}{\Delta s}.
\label{eq:lastIncre}
\end{equation} 
Substituting this term into Eq.~\eqref{eq:discretisedLapP2} and taking the continuum limit $\Delta s \to 0$, we obtain: 
\begin{equation}
\Average{\delta(y-X(s))\derpar{}{s}e^{-\int_0^s \eta(r)\left[\lambda - i\,p\,U(X(r))\right]\diff{r}}}= 
-\Phi(\lambda -i\,p\,U(y))\Average{\delta(y-X(s))e^{-\int_0^s \eta(r)\left[\lambda - i\,p\,U(X(r))\right]\diff{r}}} ,
\end{equation}   
leading with Eq.~\eqref{eq:LapPtimeDep} to the same relation Eq.~\eqref{eq:LapP2} also in the case of both space- and time-dependent forces. 
The rest of the derivation follows as in the time independent case. 
Thus, we have shown that Eq.~\eqref{eq:FFKE} with the substitution $\mathcal{L}_{\rm FP}(y) \to \mathcal{L}_{\rm FP}(y,t)=-\derpar{}{y}F(y,t)+\frac{1}{2}\dersecpar{}{y}\sigma^2(y)$ is the generalized FK formula of processes described by the subordinated Langevin Eqs.~(\ref{eq:subLE}), where external forces are allowed to depend on time, as well as on space.  
Thus, our formalism naturally provides a solution to the issue of the position of the FP operator with respect to the memory integral in a more general framework than CTRWs, in which case it has long been debated \cite{heinsalu2007use,magdziarz2008equivalence,weron2008modeling,henry2010fractional}.

\subsection{\label{Sec:4.3} Special Cases and Extensions}

Our proposed FK Equation~\eqref{eq:FFKE} recovers several different equations earlier derived in the literature for specific choices of the waiting time distribution and/or of the function $U$. 
We summarize these special cases below.

\begin{enumerate}
\item{\textit{The generalized Fokker-Planck Equation}. 
If we set $p=0$, we find a generalized Fokker-Planck  equation for the position PDF $f_Y$ \cite{magdziarz2009langevin}: 
\begin{equation}
\derpar{}{t}f_Y(y,t)=\mathcal{L}_{\rm FP}(y,t)\derpar{}{t}\int_0^{t}\,K(t-\tau)\,f_Y(y,\tau)\diff{\tau}. 
\label{eq:genFPE}
\end{equation}  
}

\item{\textit{The generalized Klein-Kramers Equation}. 
If we set $U(x)=x$ in Eq.~\eqref{eq:Wfunc}, $Y$ and $W$ correspond respectively to the velocity and the position of an anomalous diffusing particle. Thus, after inverse Fourier transform, Eq.~\eqref{eq:FFKE} yields a generalized fractional Klein-Kramers equation, which extends the result of \cite{Friedrich2006Anomalous,Friedrich2006Exact}: 
\begin{align}
\derpar{}{t}P(w,y,t)&=-\derpar{}{w}y\, P(w,y,t) +\mathcal{L}_{\rm FP}(y,t)
\left[\derpar{}{t}+\derpar{}{w}\,y \right]\!\int_0^{t}\! K(t-s) P(w-y(t-s),y,s)\diff{s}.  
\label{eq:genKKE}
\end{align}
The shift of the position sample variable in the memory integral elucidates the presence of the same retardation effects of \cite{Friedrich2006Anomalous,Friedrich2006Exact}. In particular, our derivation highlights that the stochastic dynamics underlying Eq.~\eqref{eq:genKKE} is given by the coupled Langevin Eqs.~(\ref{eq:subLE}), which has been conjectured without proof in \cite{eule2007langevin}. 
}

\item{\textit{Normal Diffusion}. 
This case is obtained with $\Phi(\lambda)=\lambda$, i.e., $K(t)=1$. In this case, Eqs.~(\ref{eq:FFKE}, \ref{eq:genFPE}, \ref{eq:genKKE}) reduce respectively to the ordinary FK , Fokker-Planck and Klein-Kramers equation \cite{risken1989fokker}. 
}

\Andrea{
\item{\textit{CTRWs with power-law waiting times}. This case is obtained by setting $\Phi(\lambda)=\lambda^\alpha$ with $0<\alpha<1$ [Eq.~\eqref{eq:LapExpStable}]. In this case, $K(t)=t^{\alpha-1}/\Gamma(\alpha)$, such that the integral operator in Eq.~(\ref{eq:FFKE}) specifies to 
\begin{equation}
\mathcal{D}^{1-\alpha}_t \widehat{P}(p,y,t)=\frac{1}{\Gamma(\alpha)}\left[ \derpar{}{t} - i\, p\, U(y) \right]\int_0^t \frac{e^{i\, p\, U(y)(t-s)}}{(t-s)^{1-\alpha}} \widehat{P}(p,y,s)\diff{s},
\end{equation}
which is the fractional substantial derivative introduced in   \cite{Friedrich2006Anomalous,Friedrich2006Exact,turgeman2009fractional,carmi2011fractional,Orzel2011FKKE}. With such choice, Eqs.~(\ref{eq:FFKE}, \ref{eq:genKKE}) become the fractional FK equation \cite{turgeman2009fractional,carmi2011fractional,Orzel2011FKKE} and the fractional Klein-Kramers equation \cite{Friedrich2006Anomalous,Friedrich2006Exact} respectively. 
If we set $p=0$, the previous operator further reduces to the Riemann-Liouville fractional derivative, i.e., Eq.~\eqref{eq:genFPE} becomes the fractional diffusion equation  \cite{metzler1999deriving,magdziarz2008equivalence}. 
}
}

\Andrea{\item{\textit{CTRWs with tempered L{\'e}vy-stable distributed waiting times}. This case is obtained by setting $\Phi(\lambda)=(\lambda+\mu)^\alpha-\mu^{\alpha}$ with $\alpha$ as in the previous case and the tempering index $\mu \in \mathbb{R}^+$ [Eq.~\eqref{eq:LapExpTempStable}]. The corresponding memory kernel in Eqs.~(\ref{eq:FFKE}, \ref{eq:genFPE}) is $K(t)=e^{-\mu t}t^{\alpha -1}E_{\alpha,\alpha}[(\mu t)^{\alpha}]$, with $E_{\alpha,\alpha}$ being a two-parameter Mittag-Leffler function. 
This specific case has also been recently discussed in \cite{wu2016tempered} by solving directly for the Laplace-Fourier transform of the joint PDF $\skew{3.5}\widehat{\widetilde{P}}(p,k,\lambda)$ of a suitable CTRW and then taking its diffusive limit, i.e., $(k,\lambda)\to(0,0)$.  
Here, we prove the equivalence of our own result and the approach therein by deriving such a limit solution. For simplicity, we restrict to time-independent external forces. Let us take the Fourier-Laplace transform of Eq.~\eqref{eq:FFKE}. Recalling that the functions $U$ and $F$ are smooth and using the convolution theorem $y f(y) \to -i \derpar{}{k}\widehat{f}(k)$, their Fourier transforms can be expressed as $U(y)\to U\!\left(-i\derpar{}{k}\right)$ and $F(y)\to F\!\left(-i\derpar{}{k}\right)$. This is understood by first Taylor expanding these functions, then Fourier transforming each term separately and finally by re-summing the series expansions. Thus, assuming the initial condition $P(w,y,t=0)=\delta(w)\delta(y)$, we obtain from Eq.~\eqref{eq:FFKE}    
\begin{align}
\lambda \skew{3.5}\widehat{\widetilde{P}}(p,k,\lambda)-1 &= i\,p\,U\!\left(-i\derpar{}{k}\right)\skew{3.5}\widehat{\widetilde{P}}(p,k,\lambda) + \left[ i\,k\,F\!\left(-i \derpar{}{k}\right)-\frac{\sigma^2}{2} k^2 \right] \frac{\lambda -i\,p\,U\!\left(-i\derpar{}{k}\right)}{\Phi\!\left(\lambda -i\,p\,U\!\left(-i\derpar{}{k}\right)\right)}\skew{3.5}\widehat{\widetilde{P}}(p,k,\lambda) . 
\end{align}
Rearranging the terms, we can rewrite it as  
\begin{align}
\left\{ \Phi\!\left(\lambda -i\,p\,U\!\left(-i\derpar{}{k}\right)\right) - \left[ i\,k\,F\!\left(-i \derpar{}{k}\right)-\frac{\sigma^2}{2} k^2 \right]\right\} \frac{\lambda -i\,p\,U\!\left(-i\derpar{}{k}\right)}{\Phi\!\left(\lambda -i\,p\,U\!\left(-i\derpar{}{k}\right)\right)} \skew{3.5}\widehat{\widetilde{P}}(p,k,\lambda) &= 1. 
\end{align}
We note that the terms in front of $\skew{3.5}\widehat{\widetilde{P}}$ are operators in the Fourier variable $k$, that do not commute in general. Therefore, applying their inverse to both sides of the previous equation in the correct order, we derive:   
\begin{align}
\skew{3.5}\widehat{\widetilde{P}}(p,k,\lambda) &= \frac{\Phi\!\left(\lambda -i\,p\,U\!\left(-i\derpar{}{k}\right)\right)}{\lambda -i\,p\,U\!\left(-i\derpar{}{k}\right)} \left\{ \Phi\!\left(\lambda -i\,p\,U\!\left(-i\derpar{}{k}\right)\right) - \left[ i\,k\,F\!\left(-i \derpar{}{k}\right)-\frac{\sigma^2}{2} k^2 \right]\right\}^{-1} .  
\label{eq:carmiF}
\end{align}
Substituting the $\Phi$ prescribed, we recover Eq.~(3) (for $F=0$) and Eq.~(16) of \cite{wu2016tempered}. Eq.~\eqref{eq:carmiF} is the formal solution in the diffusive limit of the joint PDF of a CTRW with waiting time distribution $\widetilde{\psi}(\lambda)=e^{-\Phi(\lambda)}$ \cite{turgeman2009fractional,carmi2011fractional}.  
} 
}

\item{ \textit{Multiplicative $X-$process with general $\alpha$-prescription}. 
We consider the set of subordinated Langevin equations:    
\begin{align}
\dot{X}(s)&=F(X(s),T(s))+\sigma(X(s))\star \xi(s), & 
\dot{T}(s)&=\eta(s),
\label{eq:subLEalphaP}
\end{align}
where $\star$ denotes a generalized $\alpha$ prescription in the definition of the stochastic integral (as in Eq.~\eqref{eq:BstochIalphaA} in Appendix~\ref{Sec:ItotoAlpha}). 
However, the resulting process $X$ is equivalently described by Eqs.~(\ref{eq:subLE}), i.e., with the ordinary It{\^o} prescription, by using the mapping given by Eqs.(\ref{eq:mappingItoAa}, \ref{eq:mappingItoAb}) (details are presented in Appendix~\ref{Sec:ItotoAlpha}). 
Thus, Eqs.~(\ref{eq:FFKE}, \ref{eq:memKernel}) still hold for the subordinated Eqs.~(\ref{eq:subLEalphaP}) with the modified FP operator:  
\begin{equation}
\mathcal{L}_{\rm FP}(y,t)=-\derpar{}{y}\left[F(y,t)+\alpha\,\sigma(y)\sigma^{\prime}(y)\right]+\frac{1}{2}\dersecpar{}{y}\sigma^2(y).
\label{eq:alphaFPop}
\end{equation}
The dynamics of $W$ in Eq.~\eqref{eq:Wfunc}, when $Y$ is obtained by subordination of a process $X$ of the type described by Eq.~\eqref{eq:subLEalphaP}, exhibits peculiar behavior, e.g. L{\'e}vy flight dynamics \cite{lubashevsky2009realization,lubashevsky2009continuous}, already in the Brownian limit, i.e., $\Phi(\lambda)=\lambda$. 
This motivates our interest in extending our generalized FK Eq.~\eqref{eq:FFKE} to such types of $X$-processes. 
}
\end{enumerate}

\subsection{\label{Sec:WtimeD} Explicit time dependence in the functional}

As highlighted previously in Sec.~\ref{Sec:1} a generic time-dependent protocol driving a system out of equilibrium leads both to a time-dependent force and to an explicit time-dependence in the functional, which defines the accumulated mechanical work. To the extent of our knowledge, such time-dependent functionals in the form of Eq.~\eqref{eq:Wfunc} have so far not been discussed in the literature of anomalous diffusive processes. We here address this issue by considering $W(t)$ given by Eq.~\eqref{eq:Wfunc}, where the dynamics of the underlying process $Y$ is represented by the subordinated Langevin Eqs.~(\ref{eq:subLE}). 
The time-dependence in $U$ does not modify the properties of $W$, i.e., it still has finite variation and continuous paths. Thus, the It{\^o} formula for the two-dimensional semimartingale $Z(t)=(Y(t),W(t))$ is given by Eq.~(\ref{eq:ItoSemiM2}) with the substitution $U(Y(\tau)) \rightarrow U(Y(\tau),\tau)$. 
Consequently, by repeating a similar calculation as that presented in Sec.~\ref{Sec:derivation}, we obtain the following equation: 
\begin{align}
\derpar{}{t}\widehat{P}(p,y,t)&=i\,p\,U(y,t)\,\widehat{P}(p,y,t) +\mathcal{L}_{\rm FP}(y,t)\derpar{}{t}\left\langle \int_{0}^{t}e^{i\,p\,W(\tau)}\delta(y-Y(\tau))\dot{S}(\tau)\diff{\tau}\right\rangle. 
\label{eq:prelimFFKWTD}
\end{align}
By using the same change of variables employed for Eq.~\eqref{eq:subW}, one can rewrite $W$ as below:
\begin{align}
W(t)&=A(S(t)), & A(s)&=\int_0^{s} U(X(\tau),T(\tau)) \eta(\tau)\diff{\tau}. 
\label{eq:subWTD}
\end{align}
To proceed, we note that 
(i) Eq.~(\ref{eq:GFFKstochI}) does not involve the explicit definition of $A$, such that the fundamental Eq.~\eqref{eq:Int1} also holds in this case, and 
(ii) the double average in Eq.~(\ref{eq:LapP}) can no longer be factorized, because of the dependence of $A$ on the process $T$.
Nevertheless, we can address this issue with an argument similar to that presented in Sec.~\ref{SubSec:time}. 
Thus, we first modify Eq.~\eqref{eq:LapP} as follows:
\begin{align}
\skew{3.5}\widehat{\widetilde{P}}(p,y,\lambda)&=\int_0^{+\infty} \Average{\eta(s)\,e^{-\lambda\,T(s)+i\,p\,A(s)}\,\delta(y-X(s))}\diff{s} \notag\\
&=\int_0^{+\infty} \Average{\left[\eta(s)\,e^{-\int_0^s \eta(r)\left[\lambda - i\,p\,U(X(r),T(r))\right]\diff{r}}\right]\delta(y-X(s))}\diff{s} \notag\\
&=-\int_0^{+\infty} \Average{\left[\frac{1}{\lambda-i\,p\,U(y,T(s))}\derpar{}{s}e^{-\int_0^s \eta(r)\left[\lambda - i\,p\,U(X(r),T(r))\right]\diff{r}}\right]\delta(y-X(s))} \diff{s} . \label{eq:LapPWtimeDep}
\end{align}   
Secondly, we manipulate the ensemble average appearing in its rhs by considering its discretised form. By employing the same notation used to derive Eq.~\eqref{eq:discretisedLapP2}, we obtain: 
\begin{widetext}
\begin{align}
&\Average{\left[\frac{1}{\lambda-i\,p\,U(y,T(s))}\derpar{}{s}e^{-\int_0^s \eta(r)\left[\lambda - i\,p\,U(X(r),T(r))\right]\diff{r}}\right]\delta(y-X(s))} \notag\\
&\quad =\lim_{n \to \infty \atop \Delta s \to 0}\left.\Average{\frac{1}{\lambda-i\,p\,U(y,T_n)}\frac{e^{-\sum_{j=1}^{n+1}\eta_j\,\left[\lambda -i\,p\,U(X_{j-1},T_{j-1})\right]\Delta s}-e^{-\sum_{j=1}^{n}\eta_j\,\left[\lambda -i\,p\,U(X_{j-1},T_{j-1})\right]\Delta s}}{\Delta s}}\right|_{X_n=y} \notag\\
&\quad =\lim_{n \to \infty \atop \Delta s \to 0}\left.\Average{\frac{1}{\lambda-i\,p\,U(y,T_n)}e^{-\sum_{j=1}^{n}\eta_j\,\left[\lambda -i\,p\,U(X_{j-1},T_{j-1})\right]\Delta s}\left[\frac{e^{-\Delta s\,\eta_{n+1} \left[ \lambda - i\,p\,U(X_n,T_n)\right]}-1}{\Delta s}\right]}\right|_{X_n=y} \notag\\
&\quad =\lim_{n \to \infty \atop \Delta s \to 0}\left.\Average{\frac{1}{\lambda-i\,p\,U(y,T_n)}e^{-\sum_{j=1}^{n}\eta_j\,\left[\lambda -i\,p\,U(X_{j-1})\right]\Delta s}\Average{\frac{e^{-\Delta s\,\eta_{n+1} \left[ \lambda - i\,p\,U(y,T_n)\right]}-1}{\Delta s}}}\right|_{X_n=y},
\label{eq:discretisedLapP2TD}
\end{align}  
\end{widetext}
where we explicitly separate the average over the last increment $\eta_{N+1}$ and that over the increments of both $\xi_j$ and $\eta_j$ for $j \leq n$. 
Differently from Eq.~\eqref{eq:discretisedLapP2}, these two averages cannot be factorized, because the RV $T_n$ depends on all the increments $\eta_j$. 
However, as $\eta_{n+1}$ is independent on $\eta_j$, the internal average can be solved by using Eq.~\eqref{eq:LevyChF}: 
\begin{equation} 
\Average{\frac{e^{-\Delta s\,\eta_{n+1} \left[ \lambda - i\,p\,U(y,T_n)\right]}-1}{\Delta s}}=\frac{e^{-\Delta s\,\Phi(\lambda - i\,p\,U(y,T_n))}-1}{\Delta s},
\label{eq:lastIncreWTD}
\end{equation} 
such that, by substituting it into Eq.~\eqref{eq:discretisedLapP2TD} and taking the continuum limit $\Delta s \to 0$, leads to the following relation: 
\Andrea{
\begin{multline}
\Average{\left[\frac{1}{\lambda-i\,p\,U(y,T(s))}\derpar{}{s}e^{-\int_0^s \eta(r)\left[\lambda - i\,p\,U(X(r),T(r))\right]\diff{r}}\right]\delta(y-X(s))}= \\
-\Average{\frac{\Phi(\lambda -i\,p\,U(y,T(s)))}{\lambda-i\,p\,U(y,T(s))}\,e^{-\int_0^s \eta(r)\left[\lambda - i\,p\,U(X(r),T(r))\right]\diff{r}}\, \delta(y-X(s))}.  
\end{multline}
}   
Finally, if we substitute it back into Eq.~\eqref{eq:LapPWtimeDep}, we obtain: 
\begin{equation}
\skew{3.5}\widehat{\widetilde{P}}(p,y,\lambda)= \int_0^{+\infty}\Average{\frac{\Phi\left[\lambda - i\,p\,U(y,T(s))\right]}{\lambda - i\,p\,U(y,T(s))}e^{-\lambda T(s) + i\,p\,A(s)}\delta(y-X(s))}\diff{s} . 
\label{eq:LapP2WTD}
\end{equation}
In order to close the evolution equation, we introduce the auxiliary function $\Omega$, which is defined in Laplace space as: 
\begin{equation}
\widetilde{\Omega}(\lambda)=\frac{\Phi(\lambda)}{\lambda} ,
\label{eq:w}
\end{equation}
such that the inverse Laplace transform \Andrea{(denoted as $\mathcal{L}^{-1}$)} of the time dependent terms of Eq.~\eqref{eq:LapP2WTD} can be written as
\begin{align}
\mathcal{L}^{-1}\left\{ e^{-\lambda T(s)}\frac{\Phi\left[\lambda - i\,p\,U(y,T(s))\right]}{\lambda - i\,p\,U(y,T(s))}\right\}(p,y,t)&= \int_0^t \delta(s^{\prime}-T(s))\, e^{i\,p\,U(y,T(s))(t-s^{\prime})}\, \Omega(t-s^{\prime}) \diff{s^{\prime}} \notag\\
&= \int_0^t \delta(s^{\prime}-T(s))\, e^{i\,p\,U(y,s^{\prime})(t-s^{\prime})}\, \Omega(t-s^{\prime}) \diff{s^{\prime}}.
\end{align}
By employing this result in Eq.~\eqref{eq:LapP2WTD}, we obtain the equation:
\begin{align}
\widehat{P}(p,y,t)&=\int_0^{+\infty} \left[ \int_0^t e^{i\,p\,U(y,s^{\prime})(t-s^{\prime})}\, \Omega(t-s^{\prime}) \Average{\delta(s^{\prime}-T(s))\,e^{i\,p\,A(s)}\,\delta(y-X(s))} \diff{s^{\prime}} \right] \diff{s} \notag\\
&=\int_0^{t} e^{i\,p\,U(y,s^{\prime})(t-s^{\prime})}\, \Omega(t-s^{\prime}) \left[ \int_0^{+\infty} \Average{\delta(s^{\prime}-T(s))\,e^{i\,p\,A(s)}\,\delta(y-X(s))} \diff{s} \right] \diff{s^{\prime}} ,
\label{eq:PDFWTD}
\end{align}
where in the second line we changed the order of integration. Remarkably, the term in square brackets is the same integral in the rhs of Eq.~\eqref{eq:Int1}. 
By using Eqs.~(\ref{eq:Int1}, \ref{eq:PDFWTD}), we obtain that Eq.~\eqref{eq:prelimFFKWTD} is equivalent to the coupled equations:  
\begin{align}
\derpar{}{t}\widehat{P}(p,y,t)&=i\,p\,U(y,t)\,\widehat{P}(p,y,t) +\mathcal{L}_{\rm FP}(y,t)\, H(p,y,t), \label{FFK_Uxta} \\
\widehat{P}(p,y,t)&=\int_0^{t} e^{i\,p\,U(y,s^{\prime})(t-s^{\prime})}\, \Omega(t-s^{\prime})\, H(p,y,s^{\prime}) \diff{s^{\prime}} . \label{FFK_Uxtb} 
\end{align}
Here, $H$ is an auxiliary function that is coupled to the joint PDF $\widehat{P}$. We note that in the time independent case $U(x,t)=U(x)$, Eq.~\eqref{FFK_Uxtb} reduces to a Laplace convolution. As a consequence, its Laplace transform factorizes, such that it can be solved explicitly for $H$, which can then be substituted in Eq.~\eqref{FFK_Uxta} to obtain a single closed equation, which recovers the generalized FK Eq.~\eqref{eq:FFKE} with a space- and time-dependent force.
 
\section{\label{Sec:Applications} Application to a non-equilibrium particle model} 

Since our framework includes a space- and time-dependent force as well as an explicit time-dependent functional, we can apply it to calculate the work fluctuations of an anomalous system driven by an arbitrary non-equilibrium protocol $q(t)$. Assuming a time-dependent potential $V(x,q(t))$, we consider the dynamics of Eqs.~(\ref{eq:subLE}) with $F(x,t)=-\partial_x V(x,q(t))$ and $W(t)=\int_0^t\D r\, \partial_q V(Y(r),q(r))\dot{q}(r)$. One of the simplest examples is that of a potential moving at constant velocity $v_0$. In this case we can set $q(t)=v_0t$ and $V(x,q(t))=V(x-q(t))$. This system represents the simplest pure out-of-equilibrium model, where a steady-state can be reached from the balance between dissipative forces, i.e., the friction of the surrounding fluid, and driving forces, i.e., the time dependent force due to the moving potential. Thus, it provides an easily solvable setup, where the applicability of fluctuation theorems for the accumulated mechanical work done by the system, in particular the so-called steady-state fluctuation theorem \cite{evans1993probability,gallavotti1995dynamical,gallavotti1995dynamical,gallavotti1995reversible,kurchan1998fluctuation,lebowitz1999gallavotti,seifert2012stochastic}, can be tested. 
A widely studied case is that of an harmonic potential $V(x,t)=\gamma\,(x-v_0\,t)^2/2$, such that we have in Eqs.~(\ref{eq:subLE})
\be
\label{FWharmonic}
F(x,t)=-\gamma(x-v_0\,t),\qquad\qquad\qquad\qquad W(t)=-\gamma\,v_0\int_0^t\D r\,(Y(r)-v_0r).
\ee
We also set $\sigma(x)=\sqrt{\sigma}$, where $\sigma$ is a positive real constant. In the limit case of normal diffusive dynamics, this model has been extensively studied both theoretically and experimentally \cite{wang2002experimental,van2003stationary,trepagnier2004experimental,taniguchi2007onsager,taniguchi2008nonequilibrium,gomez2010steady,aquino2013power,mestres2014realization}. The work fluctuations in the steady-state regime are described by the large deviation function $I(w)=-\lim_{t\to\infty}t^{-1}P(w,t)$, which has been calculated explicitly not only in the normal diffusive regime, but also when the random force exerted by the external bath is described by either a L{\'e}vy or a Poisson shot noise   \cite{touchette2007fluctuation,touchette2009anomalous,baule2009steady}. The solution of this paradigmatic model for anomalous dynamics of the type described in Sec.~\ref{Sec:subLangevin} has so far not been obtained.

In the anomalous case, the Eqs.~(\ref{FFK_Uxta}, \ref{FFK_Uxtb}) can be applied in principle to calculate the joint PDF $\widehat{P}(p,y,t)$ [Eq.~\eqref{ftilde}], and consequently the large deviation function $I(w)$. Due to the linear form of $W(t)$ in Eq.~\eqref{FWharmonic}(right), we can also use the simpler FK Eq.~\eqref{eq:FFKE}, since the time-dependence in the functional can be separated as below: 
\begin{equation}
W(t)=-\gamma\,v_0\,Q(t)+\gamma\,v_0^2\,\frac{t^2}{2}, \qquad\qquad\qquad\qquad Q(t)=\int_0^t\D r\,Y(r).
\label{eq:work}
\end{equation}
However, even for the linear dynamics of the dragged harmonic potential, arguably one of the simplest ways to impose a space- and time-dependent non-equilibrium drive, the linear functional case could not be fully solved so far.

\begin{figure}
\centering
\includegraphics[scale=0.3]{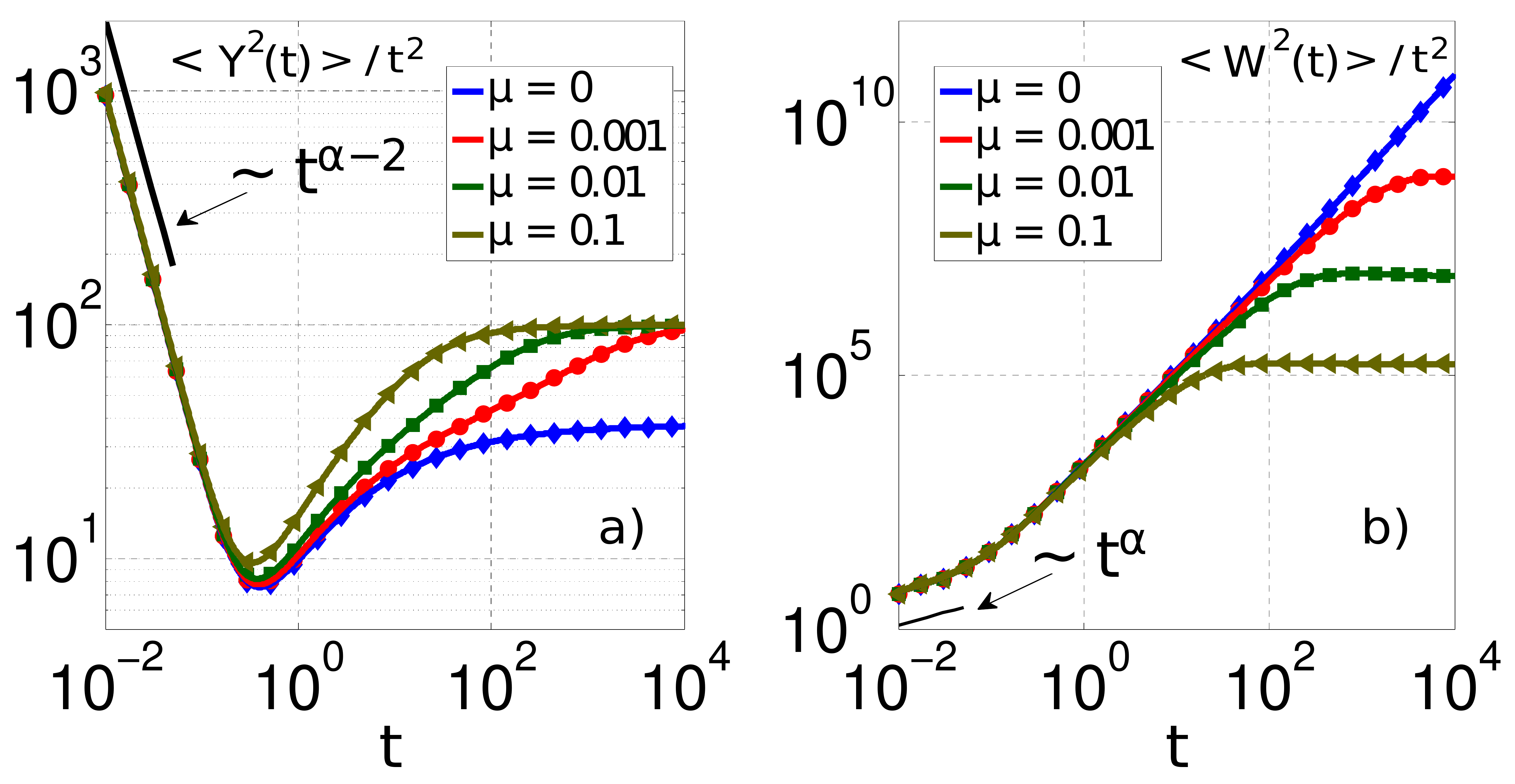}
\caption{Second-order moment of the position $Y(t)$ (a) and of the accumulated mechanical work $W(t)$ (b) of an anomalous particle driven by an harmonic potential dragged with constant velocity $v_0$. Its dynamics is described by the subordinated Langevin Eqs.~(\ref{eq:subLE}), where we choose $F(x,t)=-\gamma\,(x-v_0\,t)$ and $\sigma(x)=\sqrt{\sigma}$ with $\gamma, \sigma \in\mathbb{R}^+$. We assume tempered L{\'e}vy stable distributed waiting times with characteristic parameters $0<\alpha<1$ (order) and $\mu\in\mathbb{R}^{+}$ (tempering), i.e. $\Phi(\lambda)=(\mu+\lambda)^{\alpha}-\mu^{\alpha}$, Eq.~\eqref{eq:LapExpTempStable}. 
$W$ is expressed in terms of a linear functional $Q$ of $Y$, Eq.~\eqref{eq:work}, such that its moments are determined by both the first and second ones of $Q$ as specified by Eq.~\eqref{eq:work2mom}.   
The analytical solutions for the moments of $Y$ and $Q$ are obtained by numerical Laplace inverse transform of Eqs.~(\ref{eq:Ymsd},\ref{eq:auxG}, \ref{eq:Wmsd}, \ref{eq:auxF}) respectively. Numerical simulations (here for the specific set of parameters $\gamma=1$, $v_0=10$, $\sigma=1$ and null initial position $y_0=0$, colored markers) confirm these analytical predictions and the generalized FK Equation~\eqref{eq:FFKE} for anomalous dynamics with space-time dependent forces.  
}\label{fig:DraggedPmsd}
\end{figure}

Here, we study the first and second moment of both the position and the work. We compute the first two moments of $Y$, which will be needed later to compute the corresponding ones of $W$. The distribution of $Y$ is given by the generalized Fokker-Planck Equation~\eqref{eq:genFPE}, once we account for the correct time-dependent Fokker-Planck operator:
\begin{equation}
\derpar{}{t} P(y,t)=\left[ \gamma\,\derpar{}{y}(y-v_0\,t)+\frac{\sigma}{2}\dersecpar{}{y} \right] \derpar{}{t} \int_0^t K(t-\tau) P(y,\tau) \diff{\tau} ,
\label{eq:DPfpeq}
\end{equation}
with the memory kernel $K$ specified by Eq.~\eqref{eq:memKernel}. The general $n$-th order moment of $Y$ can be computed with the following standard procedure: (i) we take the Laplace transform of Eq.~\eqref{eq:DPfpeq} with the remark that the linear time dependent term produces a derivative in the Laplace variable; (ii) we multiply both its sides by $y^n$ (clearly $n=1,2$ for the first and second moment respectively); (iii) we perform the ensemble average, i.e., we integrate in $y$ both sides of the resulting equation. 
We remark that in Laplace space one does not need to specify beforehand the waiting time distribution, i.e., we can derive results for general $\Phi$. 
For the first moment, we obtain the following formula: 
\begin{equation}
\Average{\widetilde{Y}(\lambda)}=y_0 \frac{\Phi(\lambda)}{\lambda [\gamma + \Phi(\lambda)]} + \gamma\,v_0\,\frac{\Phi^{\prime}(\lambda)}{\lambda\,\Phi(\lambda)[\gamma + \Phi(\lambda)]} . 
\label{eq:1stXmom}
\end{equation}
In the case of CTRWs, i.e., $\Phi(\lambda)=\lambda^{\alpha}$, Eq.~\eqref{eq:1stXmom} reduces to 
$\langle \widetilde{Y}(\lambda) \rangle=\left[ y_0 + \gamma\,v_0\,\alpha\,\lambda^{-1-\alpha} \right]/[\lambda + \gamma\,\lambda^{1-\alpha}]$, whose inverse Laplace transform can be computed analytically: 
$\langle Y(t) \rangle=y_0\,\ML{\alpha}{-\gamma\,t^{\alpha}}+v_0\,\alpha\,t\,[1-\twoML{\alpha}{2}{-\gamma\,t^{\alpha}}]$. 
Furthermore, if we set $\alpha=1$ (Brownian limit) and recall that in such limit $t\,\twoML{\alpha}{2}{-\gamma\,t^{\alpha}} \rightarrow (1-e^{-\gamma\,t})/\gamma$, we obtain the expected result: 
$\langle Y(t) \rangle=y_0\, e^{-\gamma\,t} + v_0\,t - [v_0 (1-e^{-\gamma\,t})]/\gamma $ \cite{risken1989fokker,gardiner1985handbook}. 
For the second order moment, we need to compute the quantity: $\partial_{\lambda} [\lambda\,\langle \widetilde{Y}(\lambda)\rangle/\Phi(\lambda)]$, which is due to the time dependent force term. By using Eq.~\eqref{eq:1stXmom}, we find: 
\begin{subequations}
\begin{align}
\Average{\widetilde{Y}^2(\lambda)}&=y^2_0 \frac{\Phi(\lambda)}{\lambda\,[2\,\gamma + \Phi(\lambda)]} 
+\frac{\sigma}{\lambda\,[2\,\gamma + \Phi(\lambda)]} 
- \frac{2\,\gamma\,v_0}{\Phi(\lambda)+2\,\gamma}\left\{\left[\frac{1}{\lambda}-\frac{\Phi^{\prime}(\lambda)}{\Phi(\lambda)} \right]\Average{\widetilde{Y}(\lambda)}+\widetilde{G}(\lambda)\right\} ,
\label{eq:Ymsd} \\
\widetilde{G}(\lambda)&=\frac{y_0}{\lambda\,[\gamma + \Phi(\lambda)]}\! \left[ \frac{\gamma\,\Phi^{\prime}(\lambda)}{\gamma + \Phi(\lambda)}-\frac{\Phi(\lambda)}{\lambda}\right] 
\!+\! \frac{\gamma\,v_0}{\lambda\,\Phi(\lambda)[\gamma + \Phi(\lambda)]} \left\{ \Phi^{\prime\prime}(\lambda)-\frac{\Phi^{\prime}(\lambda)}{\lambda}-\frac{\gamma + 2\,\Phi(\lambda)}{\Phi(\lambda)[\gamma + \Phi(\lambda)]}[\Phi^{\prime}(\lambda)]^2 \right\} .
\label{eq:auxG}
\end{align}
\end{subequations}
In the L\'evy-stable case, $\Phi(\lambda)=\lambda^\alpha$, Eqs.~(\ref{eq:Ymsd}, \ref{eq:auxG}) can be shown to reduce to the following ones: 
\Andrea{
\begin{subequations}
\begin{align}
\Average{\widetilde{Y}^2(\lambda)}&= \frac{1}{\lambda + 2\,\gamma \lambda^{1-\alpha}} \left[ y^2_0 + 2\,\gamma\,v_0\,\widetilde{R}(\lambda) + \frac{2\,\sigma}{\lambda^{\alpha}} \right] ,
\label{eq:CTRWYmsd} \\
\widetilde{R}(\lambda)&=y_0\,\frac{\alpha\,\lambda^{\alpha-1}}{(\gamma + \lambda^{\alpha})^2}
+ \frac{\gamma\,v_0\,\alpha}{\lambda^2\,(\gamma + \lambda^{\alpha})} \left[ \frac{\alpha}{\gamma + \lambda^{\alpha}}+\frac{1+\alpha}{\lambda^{\alpha}}\right] ,
\label{eq:CTRWauxG}
\end{align}
\end{subequations}
}
which can be Laplace inverse transformed analytically as below: 
\begin{multline}
\Average{Y^2(t)}=y_0^2\,\ML{\alpha}{-2\,\gamma\,t^{\alpha}}+\frac{2\,\sigma}{\gamma} [1-\ML{\alpha}{-2\,\gamma\,t^{\alpha}}] + 2\,\gamma\,v_0^2\,\alpha(1+\alpha)\,t^{\alpha+2}\twoML{\alpha}{3+\alpha}{-2\,\gamma\,t^{\alpha}} \\
+2\,v_0\,\int_0^t \ML{\alpha}{-2\,\gamma\,(t-s)^{\alpha}}\left[ y_0\,\gamma\,s^{\alpha}\,\twoML{\alpha}{\alpha}{-\gamma\,s^{\alpha}}-\alpha\,v_0\,s\,(1-\ML{\alpha}{-\gamma\,s^{\alpha}})\right]\diff{s} .
\label{eq:dpY2moment}
\end{multline}
As a sanity check, by setting $\alpha=1$ in Eq.~\eqref{eq:dpY2moment} we obtain: $\langle Y^2(t) \rangle=y_0^2\,e^{-2\,\gamma\,t} + v_0^2\,[(1-e^{-\gamma\,t})^2/\gamma^2] + v_0^2\,t^2 + 2\,v_0\,y_0\,t\,e^{-\gamma\,t}-2\,v_0\,y_0\,e^{-\gamma\,t}\,[(1-e^{-\gamma\,t})/\gamma] -2\,v_0^2\,t\,[(1-e^{-\gamma\,t})/\gamma] + 2\,\sigma\,[(1-e^{-2\,\gamma\,t})/\gamma]$, which is the expected Brownian limit.   
In the case of the mechanical work, the first two moments are obtained by using Eq.~\eqref{eq:work}. Thus, we have: 
\begin{subequations}
\begin{align}
\langle W(t) \rangle&=-\gamma\,v_0\,\langle Q(t) \rangle + \gamma\,v_0^2\,t^2/2, \label{eq:work1mom} \\
\Average{W^2(t)}&=\gamma^2\,v_0^2\,\Average{Q^2(t)}-\gamma^2\,v_0^3\,t^2\,\Average{Q(t)}+\gamma^2\,v_0^4\,t^4/4, \label{eq:work2mom}
\end{align}
\end{subequations}
where both $\langle Q(t) \rangle$ and $\langle Q^2(t) \rangle$ can be computed analytically. On the one hand, thanks to the linearity of the functional, we find: $\langle \widetilde{Q}(\lambda) \rangle=\langle \widetilde{Y}(\lambda) \rangle/\lambda$, such that Eq.~\eqref{eq:1stXmom} can be employed to derive a closed analytic expression. On the other hand, the second order moment of $Q$ is derived by exploiting the following FK equation (Eq.~\eqref{eq:FFKE} adapted explicitly to the case considered here): 
\begin{equation}
\derpar{}{t}\widehat{P}(p,y,t)=i\,p\,y\,\widehat{P}(p,y,t) + \left[ \gamma\,\derpar{}{y} (y-v_0\,t)+\frac{\sigma}{2} \dersecpar{}{y}  \right]
\left[\derpar{}{t}-i\,p\,y\,\right]\int_0^{t}\,K(t-\tau)\,e^{i\,p\,y\,(t-\tau)}\,\widehat{P}(p,y,\tau)\diff{\tau}. \label{eq:dpFFKE} 
\end{equation}
To this aim, we need a procedure to compute joint moments of the type $\langle [Y(t)]^{m}\,[Q(t)]^{n}\rangle$. Recalling that the Fourier transform of the joint PDF of $Q$ and $Y$ is equal to $\widehat{P}(p,y,t)=\langle e^{i\,p\,Q(t)}\,\delta(y-Y(t))\rangle$, the general $n$-th order moment of $Q$ is given by  
$\langle [Q(t)]^n \rangle=\int_{-\infty}^{+\infty}\int_{-\infty}^{+\infty} a^n P(a,y,t) \diff{a}\diff{y}=(-i\,\partial_p)^{n}\int_{-\infty}^{+\infty} \widehat{P}(p,y,t) \diff{y}|_{p=0}$.
In the case of the joint moment, one simply needs to include a factor $y^m$ in the integral, i.e., 
$\langle [Y(t)]^m[Q(t)]^n \rangle=\int_{-\infty}^{+\infty}\int_{-\infty}^{+\infty} a^n\,y^m P(a,y,t) \diff{a}\diff{y}=(-i\,\partial_p)^{n}\int_{-\infty}^{+\infty} y^m\, \widehat{P}(p,y,t) \diff{y}|_{p=0}$.
Thus, the strategy to compute such moments is the following: 
(i) we take the Laplace transform of Eq.~\eqref{eq:dpFFKE} in order to express the integral term by means of $\Phi$; 
(ii) we multiply each side of the resulting equation by the corresponding power of $y$ and take its integral; 
(iii) we make the corresponding derivative in $p$ and evaluate the expression for $p=0$. 
In the specific case of the second order moment of $Q$, we obtain: 
\begin{align}
\Average{\widetilde{Q}^2(\lambda)}&=\frac{2\,\Phi(\lambda)}{\lambda^2\,[\gamma +\Phi(\lambda)]}
\left\{ \left[1+\gamma\,\widetilde{f}_2(\lambda)\right]\Average{\widetilde{Y}^2(\lambda)}+\gamma\,v_0\,\left[ \widetilde{f}_1(\lambda)\Average{\widetilde{Y}(\lambda)} -\widetilde{f}_2(\lambda)\Average{\widetilde{A}(\lambda)}-\frac{\Phi^{\prime}(\lambda)}{[\Phi(\lambda)]^2}\lambda\,\widetilde{G}(\lambda) \right]\right\} ,
\label{eq:Wmsd}
\end{align}
where we define the following auxiliary functions: 
\begin{align}
\widetilde{f}_1(\lambda)&=\frac{1}{\Phi(\lambda)}\left[\frac{1}{\lambda} - \lambda\,\frac{\Phi^{\prime\prime}(\lambda)}{\Phi(\lambda)} - \frac{2\,\Phi^{\prime}(\lambda)}{{\Phi(\lambda)}}\left(1-\lambda\,\frac{\Phi^{\prime}(\lambda)}{\Phi(\lambda)}\right)\right], \qquad\qquad\qquad
\widetilde{f}_2(\lambda)=\frac{\lambda}{\Phi(\lambda)}\left[ \frac{1}{\lambda} - \frac{\Phi^{\prime}(\lambda)}{\Phi(\lambda)}\right] .
\label{eq:auxF}
\end{align}
Our analytic results for the $Y$ and $W$ moments are in full agreement with numerical simulations, as shown in Fig.~\ref{fig:DraggedPmsd}, \Andrea{where we choose the waiting time process to be tempered L{\'e}vy-stable} [$\Phi$ as in Eq.~\eqref{eq:LapExpTempStable}]. The normal diffusive regime corresponds to a plateau in both Fig.~\ref{fig:DraggedPmsd}a and \ref{fig:DraggedPmsd}b, since both moments are plotted rescaled by $1/t^2$ and become ballistic for large times \cite{van2003stationary}. Interestingly, we observe a crossover scaling for all values of $\mu$ including the pure L\'evy-stable regime \Andrea{($\mu=0$)}. For short times, one generally observes the subdiffusive scaling $\langle Y^2(t)\rangle\sim t^{\alpha}$ for the position coordinate and the superdiffusive scaling $\langle W^2(t)\rangle\sim t^{\alpha+2}$ for the work. For long times $\langle Y^2(t)\rangle$ converges to the ballistic scaling of the normal diffusive scenario for all $\mu$, though with different ballistic diffusion coefficient between the CTRW case $\mu = 0$ and that of finite $\mu$. The behavior of $W$ is instead qualitatively different, as we find \Andrea{$\langle W^2(t)\rangle \sim t^4$} for $\mu=0$ and $\langle W^2(t)\rangle \sim t^2$ for finite $\mu$. Different intermediate scaling are observed depending on $\mu$.

\section{\label{Sec:conclusions}Conclusions and open questions} 

Despite the widespread occurrence of anomalous diffusive processes in physical, chemical and biological systems, only recently the properties of their general observables have been investigated through the systematic derivation of a FK-type equation. In this paper we demonstrated how such an equation can be derived from the stochastic description of a CTRW with generalized waiting times in the diffusive limit. Thus, our results extend the correspondence between Langevin equations and deterministic \Andrea{partial differential equations} of the original Feynman-Kac theorem to the anomalous regime. This has been \Andrea{obtained} by expressing the waiting time statistics in terms of a general one-sided L\'evy process, whose Laplace exponent turns out to be directly related to the memory kernel of the generalized FK equation. Formally, the CTRW with generalized waiting times is expressed as a normal diffusive process, subordinated by the inverse of the L\'evy process, which is itself \Andrea{generally} not a L\'evy process, but a semimartingale. \Andrea{Consequently,} our derivation requires a basic knowledge of L\'evy processes, semimartingales and their stochastic calculus, which we \Andrea{here provided} in a pedagogical way.  

While the case of space- and time-dependent forces is a non-trivial extension of our previous results in Ref.~\cite{cairoli2015anomalous}, it is manifest only in a modified Fokker-Planck operator in the generalized FK equation as expected. On the other hand, the case of an explicit time dependence in the functional leads to a new class of coupled integro-differential evolution equations for the joint PDF in Fourier space. The main challenge now is to develop \Andrea{methods to derive analytical solutions of such equations,} that could be applied to physically relevant situations. In fact, even for the simplest scenario of time independent forces and functionals, explicit solutions are sparse and have been restricted to moments or asymptotic expressions for the simplest observables with underlying L\'evy-stable \cite{Friedrich2006Anomalous,Friedrich2006Exact,turgeman2009fractional,carmi2011fractional,carmi2010distributions} and tempered L\'evy-stable waiting time processes \cite{cairoli2015anomalous,wu2016tempered}.

Open questions on a conceptual level concern the derivation of backward \Andrea{FK-type} equations in our framework and the inclusion of a time dependence in the multiplicative diffusion term. In the master equation approach, backward FK equations, i.e., where the spatial derivatives act on the space coordinate at the initial time, are straightforward to derive and are particularly relevant for occupation time problems \cite{turgeman2009fractional,carmi2011fractional,carmi2010distributions,wu2016tempered}. \Andrea{On the contrary,} in the subordination framework, the backward equations are much more challenging to treat. \Andrea{Conversely,} the case of a time- dependent diffusion term can be treated along the lines presented here, but details are left for future work.

Our results are in particular applicable to the stochastic thermodynamics of anomalous processes. Previous studies of fluctuation theorems in anomalous subdiffusive systems focused on work induced by a constant force such that the work statistics are equivalent to that of the spatial coordinate itself \cite{chechkin2009fluctuation,chechkin2012normal,dieterich2015fluctuation}. On the other hand, the mechanical work imposed by a non-equilibrium driving in a generic situation is naturally captured by our framework. A detailed discussion of work fluctuations and the associated fluctuation theorems relies on a knowledge of the large deviation function, which is currently out of reach already for the simple model of an anomalous particle in a moving harmonic potential here discussed. A further study of this paradigmatic model is certainly valuable to gain fundamental insight into the interplay of non-equilibrium driving and complex waiting time processes. Moreover, such a system can be implemented in a straightforward way in experiments, e.g., by immersing a tracer particle in a complex fluid environment and dragging it with optical tweezers \cite{tassieri2016microrheology}. \Andrea{Thus, our results pave the way for} the theoretical investigation of non-equilibrium processes of this type.


\appendix 

\section{\label{Sec:ItotoAlpha} Relation between generalized $\bm{\alpha}$ and It{\^o} Prescription}

We review the definition of the stochastic integral with respect to a Brownian motion $B(t)$. 
Let us introduce (i) a process $Y$ \Andrea{a.s.} continuous, (ii) a Brownian motion $B$ on the time interval $[0,t]$ and (iii) a partition $\pi=\{0=t_0<t_1<\ldots<t_N=t\}$ of the interval $[0,t]$ with finite mesh $|\pi|$, such that $|\pi| \to 0$. 
Thus, we can define the stochastic integral of $Y$ with respect to the increments of $B$ the following stochastic process \cite{revuz1999continuous,applebaum2009levy,karatzas2012brownian}: 
\begin{equation}
\int_0^t Y(\tau)\cdot\diff{B(\tau)}\!\!=\!\!\lim_{|\pi|\to 0 \atop N\to\infty} \sum_{i=0}^{N-1} Y(t_{i})[B(t_{i+1})-B(t_{i})]. 
\label{eq:BstochI}
\end{equation}
However, the choice of the specific time at which we evaluate the integrand process $Y$ in Eq.~\eqref{eq:BstochI} is arbitrarily chosen. 
There, this is the earlier time $t_{i}$.  
This specific choice is called the It{\^o} prescription, but in general one can choose any point in the interval $[t_{i}, t_{i+1}]$. Each of these different choices generate integrals with completely different properties. 
A general definition of the stochastic integral accounting for all the different prescriptions is given in terms of a parameter $\alpha \in [0,1]$ as follows \cite{lau2007state,lubashevsky2009realization}:       
\begin{subequations}
\begin{align}
\int_0^t Y(\tau)\star\diff{B(\tau)}&=\lim_{|\pi|\to 0 \atop N\to\infty} \sum_{i=0}^{N-1} Z(t_i) [B(t_{i+1})-B(t_{i})], \label{eq:BstochIalphaA} \\
Z(t_i)&=\left[(1-\alpha)Y(t_{i})+\alpha\, Y(t_{i+1})\right]. \label{eq:BstochIalphaB}
\end{align}
\end{subequations}
For $\alpha=0$, we recover the It{\^o} prescription, whereas for $\alpha=1/2$  we obtain the Stratonovich prescription. The case $\alpha=1$ has also been discussed in \cite{hanggi1982stochastic,klimontovich1990ito}. 
Processes of this type will be denoted with $\star$.  

We now show that 1D stochastic processes with general $\alpha$-prescription can be mapped into It{\^o} processes by suitably choosing the coefficients of the Langevin equation. 
Specifically, we consider a process $Y(t)$ described by
\begin{align}
\dot{Y}(t)&=F(t,Y(t))+\sigma(t,Y(t))\star\xi(t), \label{eq:LEPalpha}\\
\dot{Y}(t)&=a(t,Y(t))+b(t,Y(t))\cdot\xi(t), \label{eq:LEPIto}
\end{align}      
where we use respectively the generalized $\alpha$-prescription ($\alpha \neq 0$) as in Eq.~\eqref{eq:BstochIalphaB} or the It{\^o} one. Our aim is to find suitable functions $a$, $b$, such that the integrated process is the same. 
We consider the integrated version of Eq.~\eqref{eq:LEPalpha}:
\begin{equation}
Y(t)-Y_0=\int_0^t F(\tau,Y(\tau))\diff{\tau} + \int_0^t \sigma(\tau,Y(\tau))\star\diff{B(\tau)}
\label{eq:Ystar}
\end{equation} 
where the stochastic integral is defined as in Eqs.~(\ref{eq:BstochIalphaA}, \ref{eq:BstochIalphaB}). 
Here, we used the relation $\diff{B(t)}=\xi(t)\diff{t}$ between the increments of a Brownian motion and the white Gaussian noise $\xi$ \cite{gardiner1985handbook}. 
Our first task is to represent this term as an It{\^o} stochastic integral. 
To this aim, let us consider the same partition $\pi$ as before and rewrite the auxiliary variable $Z$ as $Z(t_i)=Y(t_i)+\alpha\, [Y(t_{i+1})-Y(t_i)]$. 
Thus, we can write: 
\begin{equation}
\int_0^t \sigma(\tau,Y(\tau))\star\diff{B(\tau)} 
=\lim_{|\pi|\to 0 \atop N\to\infty} \sum_{i=0}^{N-1} \sigma(t_i,Z(t_i))[B(t_{i+1})-B(t_i)]. 
\label{eq:alphaToItoI}
\end{equation} 
We note that $Z(t_i)$ depends on $\Delta Y(t_i)=Y(t_{i+1})-Y(t_i)$, which can be expressed as an It{\^o} increment by using the discretised version of Eq.~\eqref{eq:LEPIto}, i.e., we find $\Delta Y(t_i)=a(t_i,Y(t_i)) \Delta t_i+b(t_i,Y(t_i)) \Delta B(t_i)$. 
To simplify the notation, we denote: $\Delta B(t_i)=[B(t_{i+1})-B(t_i)]$ and $\Delta t_{i}=(t_{i+1}-t_i)$. Thus, we can employ such relation to express $\sigma$ as
\begin{align}
\sigma(t_i,Z(t_i))&=\sigma(t_i,Y(t_i))+\alpha\,\sigma^{\prime}(t_i,Y(t_i)) \Delta Y(t_i)
+\frac{\alpha^2}{2} \sigma(t_i,Y(t_i)) [b(t_i,Y(t_i))]^2 \Delta t_i \notag\\
&=\sigma(t_i,Y(t_i))\!+\!\alpha\, \sigma^{\prime}(t_i,Y(t_i)) b(t_i,Y(t_i)) \Delta B(t_i) \notag\\
& \qquad +\left[ \alpha\,a(t_i,Y(t_i))\sigma^{\prime}(t_i,Y(t_i)) +\frac{\alpha^2}{2} \sigma^{\prime\prime}(t_i,Y(t_i)) [b(t_i,Y(t_i))]^2 \right] \Delta t_i  
\end{align}
This result needs to be substituted back into Eq.~\eqref{eq:alphaToItoI}. 
We can then further simplify such expression by recalling that $\Average{\Delta B(t_i)^2}=\Delta t_i$,  as they are Gaussian distributed by definition, and that the $\Delta t_i$ dependent term cancels out in the limit of null mesh. Thus we obtain:     
\begin{align}
&\int_0^t \sigma(\tau,Y(\tau))\star\diff{B(\tau)}=\int_0^t \sigma(\tau,Y(\tau))\cdot\diff{B}(\tau) + \alpha \int_0^t \sigma^{\prime}(\tau,Y(\tau)) b(\tau ,Y(\tau))\diff{\tau}. 
\label{eq:alphaToIto}
\end{align} 
By using Eqs.~(\ref{eq:Ystar}, \ref{eq:alphaToIto}) together, we obtain: 
\begin{align}
Y(t)-Y_0&=\int_0^t \left[ F(\tau,Y(\tau))+ \alpha\,\sigma^{\prime}(\tau,Y(\tau))b(\tau,Y(\tau))\right] \diff{\tau} + \int_0^t \sigma(\tau,Y(\tau))\cdot\diff{B(\tau)}.  
\end{align} 
It is now clear that the mapping between the two processes is realised if we set: 
\begin{subequations}
\begin{align}
b(t,Y(t))&\!=\!\sigma(t,Y(t)), \label{eq:mappingItoAa} \\
a(t,Y(t))&\!=\!F(t,Y(t))+\alpha\,\sigma(t,Y(t))\sigma^{\prime}(t,Y(t)). 
\label{eq:mappingItoAb}
\end{align}
\end{subequations}

\section{\label{Sec:FVP} Finite variation processes}

We review definition and properties of continuous stochastic processes with paths of finite variation. Specifically, we will provide the definition of their stochastic integral and their It{\^o} formula. As the time-change process $S$ defined in Sec.~\ref{Sec:subLangevin} belongs to this class, such notions are employed in the derivation of the generalized fractional FK Eq.~\eqref{eq:FFKE}.   
     
As a preliminary step, we define the total variation of a real-valued function $g$ with support on an interval $[s,t]$. 
Thus, we consider a partition of the interval $\pi=\{s=t_0<t_1<\ldots<t_N=t\}$, whose mesh is given by the maximum of the lengths of the subintervals: $|\pi|=\max_{i=1,\ldots,N} |t_{i}-t_{i-1}|$ and compute the quantity:  
\begin{equation}
V_t^{\pi}(g)=\sum_{i=1}^N |g(t_i)-g(t_{i-1})|,  
\label{eq:1stVar}
\end{equation}
whose value clearly depends on the specific $\pi$ chosen. 
Let us now consider the set of all possible partitions $\mathcal{P}=\{\pi_n\}$ and the corresponding variations of $g$ with respect to them $\{V_t^{\pi_n}(g)\}$. The total variation of $g$ on $[s,t]$ is obtained by taking the supremum of this set: 
\begin{equation}
V_t(g)=\sup_{\pi \in \mathcal{P}} V_t^{\pi}(g). 
\end{equation}
Thus, if $V_t(g)<\infty$, then $g$ is said to be of finite variation and $V_t(g)$ is the total variation of $g$ on the chosen interval; otherwise, it is said to have infinite variation. 
If $g$ is defined over all $\mathbb{R}$, then $g$ has finite variation if it is of finite variation on all closed intervals of $\mathbb{R}$. Clearly, if $g$ is a non decreasing function, then it is of finite variation, as $V_t(g)=g(t_N)-g(t_0)$. Conversely, if $g$ is of finite variation, we can always find two auxiliary non decreasing functions $g_1$ and $g_2$, such that $g=g_1+g_2$.   

In a similar way, a stochastic process $Y$ is said to be of finite variation if its stochastic trajectories $Y(t)$ have finite variation almost surely, i.e., for each of its different realizations. An analogous definition holds in the opposite case of a process of infinite variation. We note that ordinary integrals (in Lebesgue sense) of a continuous stochastic process are also of finite variation.  

Stochastic integrals with respect to these processes can be defined straightforwardly as Lebesgue-Stieltjes integral with the proper measure associated to $Y$, which exists due to the finite variation of their paths \cite{ash2000probability}. 
In terms of Riemann sums, considering the same partition $\pi$ as before, the stochastic integral of an arbitrary function $H(t)$ with respect to $Y$ is defined as
\begin{equation}
\int_0^t H(\tau)\diff{Y(\tau)}\!=\!\lim_{N \to \infty \atop |\pi|\to 0} \sum_{i=1}^{N} H(t_i)[Y(t_i)-Y(t_{i-1})].
\label{eq:stochIFVP}
\end{equation} 
We note that $H$ does not need to be continuous, but its paths are required to be right continuous with left limits (\textit{c{\`a}dl{\`a}g}). We further remark that the stochastic integral in Eq.~\eqref{eq:stochIFVP} can also be defined when $Y$ has general c{\`a}dl{\`a}g paths, i.e., not continuous. However, in such case jump terms need to be properly accounted for. 
As this is not the case of $S$, we will not discuss it in this context. 
For a general differentiable function $f$ of $Y$ the It{\^o} formula is
\begin{equation}
f(Y(t))=f(Y_0)+\int_0^t f^{\prime}(Y(\tau))\diff{Y(\tau)}.
\label{eq:ItoFVP}
\end{equation}
This follows straightforwardly by considering again the partition $\pi$ and employing the mean value theorem:   
\begin{align}
f(Y(t))-f(Y_0)&=\sum_{i=1}^{N} [f(Y(t_i))-f(Y(t_{i-1})]\notag\\
&=\sum_{i=1}^{N} f^{\prime}(Y(t_i))(t_i-t_{i-1}) 
=\int_0^t f^{\prime}(Y(\tau))\diff{\tau} . 
\end{align}

\section{\label{Sec:QuadVar} The quadratic variation}

We here define the quadratic variation of a process $Y(t)$ for $t \geq 0$ on a time interval $[s,t]$.  Let us consider a partition of such interval $\pi=\{s=t_0<t_1<\ldots<t_N=t\}$ of mesh $|\pi|=\max_{i=1,\ldots,N} |t_{i}-t_{i-1}|$. Associated to $\pi$, we can define the following process: 
\begin{equation}
[Y,Y]^{\pi}_t=\sum_{i=1}^N [Y(t_i)-Y(t_{i-1})]^2 .
\label{eq:quadVar}
\end{equation}
Clearly, $[Y,Y]^{\pi}_t$ depends both on the specific realization of $Y$ and on the partition chosen. To avoid this latter dependence, we study its properties in the limit $|\pi| \to 0$. Let us now consider sequences of partitions $\{\pi_n\}$, such that $|\pi_n|\to 0$, and compute the corresponding sequences $\{[Y,Y]^{\pi_n}_t\}$. If for every $t$ this latter sequence converges in probability to a finite value $[Y,Y]_t$ independent on the specific choice of $\{\pi_n\}$ a.s., then $[Y,Y]_t$ is a well-defined process called the quadratic variation of $Y$. 
As an example, we show that the quadratic variation of a process $Y$ with paths of finite variation exists and it is null. From Eq.~\eqref{eq:quadVar} and for a given $\pi$, we can write:  
\begin{align}
[Y,Y]^{\pi}_t & \leq \left( \sum_{i=1}^N |Y(t_i)-Y(t_{i-1})|\right)\, \max_{j=1, \ldots, N}|Y(t_j)-Y(t_{j-1})| \notag\\
& \leq V_t^{\pi}(Y)\, \max_{j=1, \ldots, N}|Y(t_j)-Y(t_{j-1})|= V_t^{\pi}(Y)\, |\pi| , 
\end{align}
where $V_t^{\pi}(Y)$ is the variation of $Y$ as defined in Eq.~\eqref{eq:quadVar}, which is finite by assumption. Thus, the rhs converges to zero a.s. in the limit $|\pi|\to 0$. As this result holds independently of the specific $\pi$, we conclude that a.s. $[Y,Y]_t=0$.


\begin{thebibliography}{100}

\bibitem{risken1989fokker}
H.~Risken.
\newblock The {F}okker-{P}lanck {E}quation. {M}ethods of {S}olution and
  {A}pplications.
\newblock {\em Springer Series in Synergetics}, 1989.

\bibitem{darling1957occupation}
D.~A. Darling and M.~Kac.
\newblock On occupation times for {M}arkoff processes.
\newblock {\em Trans. Amer. Math. Soc.}, 84(2):444--458, 1957.

\bibitem{agmon1984residence}
N.~Agmon.
\newblock Residence times in diffusion processes.
\newblock {\em J. Chem. Phys.}, 81(8):3644--3647, 1984.

\bibitem{newman1998diffusive}
T.~J. Newman and Z.~Toroczkai.
\newblock Diffusive persistence and the ``sign-time'' distribution.
\newblock {\em Phys. Rev. E}, 58:R2685--R2688, Sep 1998.

\bibitem{berezhkovskii1998residence}
A.~M. Berezhkovskii, V.~Zaloj, and N.~Agmon.
\newblock Residence time distribution of a {B}rownian particle.
\newblock {\em Phys. Rev. E}, 57(4):3937, 1998.

\bibitem{dhar1999residence}
A.~Dhar and S.~N. Majumdar.
\newblock Residence time distribution for a class of {G}aussian {M}arkov
  processes.
\newblock {\em Phys. Rev. E}, 59(6):6413, 1999.

\bibitem{godreche2001statistics}
C.~Godreche and J.~M. Luck.
\newblock Statistics of the occupation time of renewal processes.
\newblock {\em J. Stat. Phys.}, 104(3-4):489--524, 2001.

\bibitem{majumdar2002local}
S.~N. Majumdar and A.~Comtet.
\newblock Local and occupation time of a particle diffusing in a random medium.
\newblock {\em Phys. Rev. Lett.}, 89(6):060601, 2002.

\bibitem{blanco2003invariance}
S.~Blanco and R.~Fournier.
\newblock An invariance property of diffusive random walks.
\newblock {\em EPL}, 61(2):168, 2003.

\bibitem{mazzolo2004properties}
A.~Mazzolo.
\newblock Properties of diffusive random walks in bounded domains.
\newblock {\em EPL}, 68(3):350, 2004.

\bibitem{barkai2006residence}
E.~Barkai.
\newblock Residence time statistics for normal and fractional diffusion in a
  force field.
\newblock {\em J. Stat. Phys.}, 123(4):883--907, 2006.

\bibitem{grebenkov2007residence}
D.~S. Grebenkov.
\newblock Residence times and other functionals of reflected {B}rownian motion.
\newblock {\em Phys. Rev. E}, 76(4):041139, 2007.

\bibitem{dumonteil2016residence}
E.~Dumonteil and A.~Mazzolo.
\newblock Residence times of branching diffusion processes.
\newblock {\em Phys. Rev. E}, 94(1):012131, 2016.

\bibitem{majumdar2005brownian}
S.~N. Majumdar.
\newblock Brownian functionals in physics and computer science.
\newblock {\em Curr. Sci.}, 88, 2005.

\bibitem{yor2012exponential}
M.~Yor.
\newblock {\em Exponential functionals of Brownian motion and related
  processes}.
\newblock Springer Science \& Business Media, 2012.

\bibitem{jarzynski1997nonequilibrium}
C.~Jarzynski.
\newblock Nonequilibrium equality for free energy differences.
\newblock {\em Phys. Rev. Lett.}, 78(14):2690, 1997.

\bibitem{sekimoto1997kinetic}
K.~Sekimoto.
\newblock Kinetic characterization of heat bath and the energetics of thermal
  ratchet models.
\newblock {\em J. Phys. Soc. Jpn.}, 66(5):1234--1237, 1997.

\bibitem{sekimoto1998langevin}
K.~Sekimoto.
\newblock Langevin equation and thermodynamics.
\newblock {\em Prog. Theor. Phys.}, 130:17--27, 1998.

\bibitem{sekimoto2010stochastic}
K.~Sekimoto.
\newblock {\em Stochastic energetics}, volume 799.
\newblock Springer, 2010.

\bibitem{seifert2012stochastic}
U.~Seifert.
\newblock Stochastic thermodynamics, fluctuation theorems and molecular
  machines.
\newblock {\em Rep. Prog. Phys.}, 75(12):126001, 2012.

\bibitem{peseckis1987statistical}
F.~E. Peseckis.
\newblock Statistical dynamics of stable processes.
\newblock {\em Phys. Rev. A}, 36(2):892, 1987.

\bibitem{fogedby1994levy}
H.~C. Fogedby.
\newblock L{\'e}vy flights in random environments.
\newblock {\em Phys. Rev. Lett.}, 73(19):2517, 1994.

\bibitem{jespersen1999levy}
S.~Jespersen, R.~Metzler, and H.~C. Fogedby.
\newblock L{\'e}vy flights in external force fields: {L}angevin and fractional
  {F}okker-{P}lanck equations and their solutions.
\newblock {\em Phys. Rev. E}, 59(3):2736, 1999.

\bibitem{lutz2001fractional}
Eric Lutz.
\newblock Fractional transport equations for {L\'e}vy stable processes.
\newblock {\em Phys. Rev. Lett.}, 86(11):2208, 2001.

\bibitem{eliazar2003levy}
I.~Eliazar and J.~Klafter.
\newblock L{\'e}vy-driven {L}angevin systems: {T}argeted stochasticity.
\newblock {\em J. Stat. Phys.}, 111(3-4):739--768, 2003.

\bibitem{metzler2000generalized1}
R.~Metzler.
\newblock Generalized {C}hapman-{K}olmogorov equation: {A} unifying approach to
  the description of anomalous transport in external fields.
\newblock {\em Phys. Rev. E}, 62(5):6233, 2000.

\bibitem{metzler2000subdiffusive}
R.~Metzler and J.~Klafter.
\newblock Subdiffusive transport close to thermal equilibrium: from the
  {L}angevin equation to fractional diffusion.
\newblock {\em Phys. Rev. E}, 61(6):6308, 2000.

\bibitem{metzler2000generalized2}
R.~Metzler and J.~Klafter.
\newblock From a generalized {C}hapman-{K}olmogorov equation to the fractional
  {K}lein-{K}ramers equation.
\newblock {\em J. Phys. Chem. B}, 104(16):3851--3857, 2000.

\bibitem{barkai2000fractional}
E.~Barkai and R.~J. Silbey.
\newblock Fractional {K}ramers equation.
\newblock {\em J. Phys. Chem. B}, 104(16):3866--3874, 2000.

\bibitem{metzler2002superdiffusive}
R.~Metzler and I.~M. Sokolov.
\newblock Superdiffusive {K}lein-{K}ramers equation: {N}ormal and anomalous
  time evolution and {L\'e}vy walk moments.
\newblock {\em EPL}, 58(4):482, 2002.

\bibitem{zoia2012discrete}
A.~Zoia, E.~Dumonteil, and A.~Mazzolo.
\newblock Discrete {F}eynman-{K}ac formulas for branching random walks.
\newblock {\em EPL}, 98(4):40012, 2012.

\bibitem{fa2013generalized}
K.~S. Fa and K.~G. Wang.
\newblock Generalized {K}lein--{K}ramers equation: solution and application.
\newblock {\em J. Stat. Mech.}, 2013(09):P09021, 2013.

\bibitem{dieterich2008anomalous}
P.~Dieterich, R.~Klages, R.~Preuss, and A.~Schwab.
\newblock Anomalous dynamics of cell migration.
\newblock {\em Proc. Natl. Acad. Sci.}, 105(2):459--463, 2008.

\bibitem{baule2006investigation}
A.~Baule and R.~Friedrich.
\newblock Investigation of a generalized {O}bukhov model for turbulence.
\newblock {\em Phys. Lett. A}, 350(3):167--173, 2006.

\bibitem{eule2007langevin}
S.~Eule, R.~Friedrich, F.~Jenko, and D.~Kleinhans.
\newblock Langevin approach to fractional diffusion equations including
  inertial effects.
\newblock {\em J. Phys. Chem. B}, 111(39):11474--11477, 2007.

\bibitem{eule2012langevin}
S.~Eule, V.~Zaburdaev, R.~Friedrich, and T.~Geisel.
\newblock Langevin description of superdiffusive {L\'e}vy processes.
\newblock {\em Phys. Rev. E}, 86(4):041134, 2012.

\bibitem{montroll1965random}
E.~W. Montroll and G.~H. Weiss.
\newblock Random walks on lattices. {II}.
\newblock {\em J. Math. Phys.}, 6:167, 1965.

\bibitem{metzler2000random}
R.~Metzler and J.~Klafter.
\newblock The random walk's guide to anomalous diffusion: a fractional dynamics
  approach.
\newblock {\em Phys. Rep.}, 339(1):1--77, 2000.

\bibitem{Friedrich2006Anomalous}
R.~Friedrich, F.~Jenko, A.~Baule, and S.~Eule.
\newblock Anomalous {D}iffusion of {I}nertial, {W}eakly {D}amped {P}articles.
\newblock {\em Phys. Rev. Lett.}, 96:230601, Jun 2006.

\bibitem{Friedrich2006Exact}
R.~Friedrich, F.~Jenko, A.~Baule, and S.~Eule.
\newblock Exact solution of a generalized {K}ramers-{F}okker-{P}lanck equation
  retaining retardation effects.
\newblock {\em Phys. Rev. E}, 74:041103, 2006.

\bibitem{turgeman2009fractional}
L.~Turgeman, S.~Carmi, and E.~Barkai.
\newblock Fractional {F}eynman-{K}ac equation for non-{B}rownian functionals.
\newblock {\em Phys. Rev. Lett.}, 103(19):190201, 2009.

\bibitem{carmi2010distributions}
S.~Carmi, L.~Turgeman, and E.~Barkai.
\newblock On distributions of functionals of anomalous diffusion paths.
\newblock {\em J. Stat. Phys.}, 141(6):1071--1092, 2010.

\bibitem{carmi2011fractional}
S.~Carmi and E.~Barkai.
\newblock Fractional {F}eynman-{K}ac equation for weak ergodicity breaking.
\newblock {\em Phys. Rev. E}, 84(6):061104, 2011.

\bibitem{shkilev2012equations}
V.~P. Shkilev.
\newblock Equations for the distributions of functionals of a random-walk
  trajectory in an inhomogeneous medium.
\newblock {\em J. Exp. Theor. Phys.}, 114(1):172--181, 2012.

\bibitem{shkilev2016feynman}
V.~P. Shkilev.
\newblock Feynman-{K}ac {E}quations for {R}andom {W}alks in {D}isordered
  {M}edia.
\newblock {\em Math. Model. Nat. Phenom.}, 11(3):63--75, 2016.

\bibitem{cairoli2015anomalous}
A.~Cairoli and A.~Baule.
\newblock Anomalous {P}rocesses with {G}eneral {W}aiting {T}imes: {F}unctionals
  and {M}ultipoint {S}tructure.
\newblock {\em Phys. Rev. Lett.}, 115(11):110601, 2015.

\bibitem{wu2016tempered}
X.~Wu, W.~Deng, and E.~Barkai.
\newblock Tempered fractional {F}eynman-{K}ac equation: {T}heory and examples.
\newblock {\em Phys. Rev. E}, 93:032151, 2016.

\bibitem{selmeczi2005cell}
D.~Selmeczi, S.~Mosler, P.~H. Hagedorn, N.~B. Larsen, and H.~Flyvbjerg.
\newblock Cell motility as persistent random motion: theories from experiments.
\newblock {\em Biophys. J.}, 89(2):912--931, 2005.

\bibitem{selmeczi2008cell}
D.~Selmeczi, L.~Li, L.~I.~I. Pedersen, S.~F. Nrrelykke, P.~H. Hagedorn,
  S.~Mosler, N.~B. Larsen, E.~C. Cox, and H.~Flyvbjerg.
\newblock Cell motility as random motion: a review.
\newblock {\em Eur. Phys. J. Spec.}, 157(1):1--15, 2008.

\bibitem{campos2010persistent}
D.~Campos, V.~M{\'e}ndez, and I.~Llopis.
\newblock Persistent random motion: {U}ncovering cell migration dynamics.
\newblock {\em J. Theor. Biol.}, 267(4):526--534, 2010.

\bibitem{harris2012generalized}
T.~H. Harris, E.~J. Banigan, D.~A. Christian, C.~Konradt, E.~D.~T Wojno,
  K.~Norose, E.~H. Wilson, B.~John, W.~Weninger, A.~D. Luster, et~al.
\newblock Generalized {L\'e}vy walks and the role of chemokines in migration of
  effector {CD}8+ {T} cells.
\newblock {\em Nature}, 486(7404):545--548, 2012.

\bibitem{caspi2000enhanced}
A.~Caspi, R.~Granek, and M.~Elbaum.
\newblock Enhanced diffusion in active intracellular transport.
\newblock {\em Phys. Rev. Lett.}, 85(26):5655, 2000.

\bibitem{levi2005chromatin}
V.~Levi, Q.~Ruan, M.~Plutz, A.~S. Belmont, and E.~Gratton.
\newblock Chromatin dynamics in interphase cells revealed by tracking in a
  two-photon excitation microscope.
\newblock {\em Biophys. J.}, 89(6):4275--4285, 2005.

\bibitem{brangwynne2007force}
C.~P. Brangwynne, F.~C. MacKintosh, and D.~A. Weitz.
\newblock Force fluctuations and polymerization dynamics of intracellular
  microtubules.
\newblock {\em Proc. Natl. Acad. Sci.}, 104(41):16128--16133, 2007.

\bibitem{bronstein2009transient}
I.~Bronstein, Y.~Israel, E.~Kepten, S.~Mai, Y.~Shav-Tal, E.~Barkai, and
  Y.~Garini.
\newblock Transient anomalous diffusion of telomeres in the nucleus of
  mammalian cells.
\newblock {\em Phys. Rev. Lett.}, 103(1):018102, 2009.

\bibitem{bruno2009transition}
L.~Bruno, V.~Levi, M.~Brunstein, and M.~A. Desp{\'o}sito.
\newblock Transition to superdiffusive behavior in intracellular actin-based
  transport mediated by molecular motors.
\newblock {\em Phys. Rev. E}, 80(1):011912, 2009.

\bibitem{senning2010actin}
E.~N. Senning and A.~H. Marcus.
\newblock Actin polymerization driven mitochondrial transport in mating {S}.
  cerevisiae.
\newblock {\em Proc. Natl. Acad. Sci.}, 107(2):721--725, 2010.

\bibitem{Jeon2011InVivo}
J.-H. Jeon, V.~Tejedor, S.~Burov, E.~Barkai, C.~Selhuber-Unkel,
  K.~Berg-S\o{}rensen, L.~Oddershede, and R.~Metzler.
\newblock \textit{In Vivo} {Anomalous Diffusion and Weak Ergodicity Breaking of
  Lipid Granules}.
\newblock {\em Phys. Rev. Lett.}, 106:048103, 2011.

\bibitem{jeon2012anomalous}
J.-H. Jeon, H.~M.-S. Monne, M.~Javanainen, and R.~Metzler.
\newblock Anomalous diffusion of phospholipids and cholesterols in a lipid
  bilayer and its origins.
\newblock {\em Phys. Rev. Lett.}, 109(18):188103, 2012.

\bibitem{weber2012nonthermal}
S.~C. Weber, A.~J. Spakowitz, and J.~A. Theriot.
\newblock Nonthermal {ATP}-dependent fluctuations contribute to the in vivo
  motion of chromosomal loci.
\newblock {\em Proc. Natl. Acad. Sci.}, 109(19):7338--7343, 2012.

\bibitem{von2013anomalous}
Y.~von Hansen, S.~Gekle, and R.~R. Netz.
\newblock Anomalous anisotropic diffusion dynamics of hydration water at lipid
  membranes.
\newblock {\em Phys. Rev. Lett.}, 111(11):118103, 2013.

\bibitem{tabei2013intracellular}
S.~M.~A. Tabei, S.~Burov, H.~Y. Kim, A.~Kuznetsov, T.~Huynh, J.~Jureller, L.~H.
  Philipson, A.~R. Dinner, and N.~F. Scherer.
\newblock Intracellular transport of insulin granules is a subordinated random
  walk.
\newblock {\em Proc. Natl. Acad. Sci.}, 110(13):4911--4916, 2013.

\bibitem{javer2014persistent}
A.~Javer, N.~J. Kuwada, Z.~Long, V.~G. Benza, K.~D. Dorfman, P.~A. Wiggins,
  P.~Cicuta, and M.~C. Lagomarsino.
\newblock Persistent super-diffusive motion of {E}scherichia coli chromosomal
  loci.
\newblock {\em Nat. Commun.}, 5, 2014.

\bibitem{tejedor2010anomalous}
V.~Tejedor and R.~Metzler.
\newblock Anomalous diffusion in correlated continuous time random walks.
\newblock {\em J. Phys. A}, 43(8):082002, 2010.

\bibitem{magdziarz2012correlated}
M.~Magdziarz, R.~Metzler, W.~Szczotka, and P.~{\.Z}ebrowski.
\newblock Correlated continuous-time random walks—scaling limits and
  {L}angevin picture.
\newblock {\em J. Stat. Mech.}, 2012(04):P04010, 2012.

\bibitem{magdziarz2012langevin}
M.~Magdziarz, W.~Szczotka, and P.~{\.Z}ebrowski.
\newblock Langevin picture of {L}evy walks and their extensions.
\newblock {\em J. Stat. Phys.}, 147(1):74--96, 2012.

\bibitem{schulz2013correlated}
J.~H.~P. Schulz, A.~V. Chechkin, and R.~Metzler.
\newblock Correlated continuous time random walks: combining scale-invariance
  with long-range memory for spatial and temporal dynamics.
\newblock {\em J. Phys. A}, 46(47):475001, 2013.

\bibitem{de2013flow}
P.~De~Anna, T.~Le~Borgne, M.~Dentz, A.~M. Tartakovsky, D.~Bolster, and P.~Davy.
\newblock Flow intermittency, dispersion, and correlated continuous time random
  walks in porous media.
\newblock {\em Phys. Rev. Lett.}, 110(18):184502, 2013.

\bibitem{liu2013continuous}
J.~Liu and J.-D. Bao.
\newblock Continuous time random walk with jump length correlated with waiting
  time.
\newblock {\em Physica A}, 392(4):612--617, 2013.

\bibitem{magdziarz2013asymptotic}
M.~Magdziarz, W.~Szczotka, and P.~{\.Z}.ebrowski.
\newblock Asymptotic behaviour of random walks with correlated temporal
  structure.
\newblock {\em Proc. R. Soc. A}, 469(2159):20130419, 2013.

\bibitem{metzler1998fractional}
R.~Metzler and T.~F. Nonnenmacher.
\newblock Fractional diffusion, waiting-time distributions, and {C}attaneo-type
  equations.
\newblock {\em Phys. Rev. E}, 57(6):6409, 1998.

\bibitem{metzler1998anomalous}
R.~Metzler, J.~Klafter, and I.~M. Sokolov.
\newblock Anomalous transport in external fields: {C}ontinuous time random
  walks and fractional diffusion equations extended.
\newblock {\em Phys. Rev. E}, 58(2):1621, 1998.

\bibitem{metzler1999deriving}
R.~Metzler, E.~Barkai, and J.~Klafter.
\newblock Deriving fractional {F}okker-{P}lanck equations from a generalised
  master equation.
\newblock {\em EPL}, 46(4):431, 1999.

\bibitem{metzler1999anomalousPRL}
R.~Metzler, E.~Barkai, and J.~Klafter.
\newblock Anomalous diffusion and relaxation close to thermal equilibrium: a
  fractional {F}okker-{P}lanck equation approach.
\newblock {\em Phys. Rev. Lett.}, 82(18):3563, 1999.

\bibitem{metzler1999transport}
R.~Metzler, E.~Barkai, and J.~Klafter.
\newblock Anomalous transport in disordered systems under the influence of
  external fields.
\newblock {\em Physica A}, 266(1):343--350, 1999.

\bibitem{barkai2000continuous}
E.~Barkai, R.~Metzler, and J.~Klafter.
\newblock From continuous time random walks to the fractional {F}okker-{P}lanck
  equation.
\newblock {\em Phys. Rev. E}, 61(1):132, 2000.

\bibitem{fogedby1994langevin}
H.~C. Fogedby.
\newblock Langevin equations for continuous time {L\'e}vy flights.
\newblock {\em Phys. Rev. E}, 50(2):1657, 1994.

\bibitem{baule2005joint}
A.~Baule and R.~Friedrich.
\newblock Joint probability distributions for a class of non-{M}arkovian
  processes.
\newblock {\em Phys. Rev. E}, 71(2):026101, 2005.

\bibitem{weron2008modeling}
A.~Weron, M.~Magdziarz, and K.~Weron.
\newblock Modeling of subdiffusion in space-time-dependent force fields beyond
  the fractional {F}okker-{P}lanck equation.
\newblock {\em Phys. Rev. E}, 77(3):036704, 2008.

\bibitem{magdziarz2007fractional}
M.~Magdziarz, A.~Weron, K.~Weron, et~al.
\newblock Fractional {Fokker-Planck} dynamics: Stochastic representation and
  computer simulation.
\newblock {\em Phys. Rev. E}, 75(1):016708, 2007.

\bibitem{magdziarz2008equivalence}
M.~Magdziarz, A.~Weron, and J.~Klafter.
\newblock Equivalence of the fractional {Fokker-Planck} and subordinated
  {Langevin} equations: the case of a time-dependent force.
\newblock {\em Phys. Rev. Lett.}, 101(21):210601, 2008.

\bibitem{henry2010fractional}
B.~I. Henry, T.~A.~M. Langlands, and P.~Straka.
\newblock Fractional {Fokker-Planck} equations for subdiffusion with space-and
  time-dependent forces.
\newblock {\em Phys. Rev. Lett.}, 105(17):170602, 2010.

\bibitem{meerschaert2011fractional}
M.~M Meerschaert, E.~Nane, P.~Vellaisamy, et~al.
\newblock The fractional poisson process and the inverse stable subordinator.
\newblock {\em Electron. J. Probab.}, 16(59):1600--1620, 2011.

\bibitem{cont1975financial}
R.~Cont and P.~Tankov.
\newblock {\em Financial {Modelling} with jump processes}.
\newblock CRC Press, London, 2003.

\bibitem{applebaum2009levy}
D.~Applebaum.
\newblock {\em L{\'e}vy processes and stochastic calculus}.
\newblock Cambridge University Press, Cambridge, 2009.

\bibitem{magdziarz2009langevin}
M.~Magdziarz.
\newblock Langevin picture of subdiffusion with infinitely divisible waiting
  times.
\newblock {\em J. Stat. Phys.}, 135, 2009.

\bibitem{sandev2015distributed}
T.~Sandev, A.V. Checkin, N.~Korabel, H.~Kantz, I.~Sokolov, and R.~Metzler.
\newblock Distributed-order diffusion equations and multifractality: {M}odels
  and solutions.
\newblock {\em Phys. Rev. E}, 92(4):042117, 2015.

\bibitem{fedotov2014sub}
S.~Fedotov and N.~Korabel.
\newblock Subdiffusion in an external potential: {Anomalous} effects hiding
  behind normal behavior.
\newblock {\em Phys. Rev. E}, 91:042112, 2015.

\bibitem{cairoli2015langevin}
A.~Cairoli and A.~Baule.
\newblock Langevin formulation of a subdiffusive continuous-time random walk in
  physical time.
\newblock {\em Phys. Rev. E}, 92:012102, Jul 2015.

\bibitem{heinsalu2007use}
E.~Heinsalu, M.~Patriarca, I.~Goychuk, and P.~H{\"a}nggi.
\newblock Use and abuse of a fractional {Fokker-Planck} dynamics for
  time-dependent driving.
\newblock {\em Phys. Rev. Lett.}, 99(12):120602, 2007.

\bibitem{magdziarz2009stochastic}
M.~Magdziarz.
\newblock Stochastic representation of subdiffusion processes with
  time-dependent drift.
\newblock {\em Stoch. Proc. Appl.}, 119(10):3238--3252, 2009.

\bibitem{eule2009subordinated}
S.~Eule and R.~Friedrich.
\newblock Subordinated langevin equations for anomalous diffusion in external
  potentials--biasing and decoupled external forces.
\newblock {\em EPL}, 86(3):30008, 2009.

\bibitem{heinsalu2009fractional}
E.~Heinsalu, M.~Patriarca, I.~Goychuk, and P.~H{\"a}nggi.
\newblock Fractional {Fokker-Planck} subdiffusion in alternating force fields.
\newblock {\em Phys. Rev. E}, 79(4):041137, 2009.

\bibitem{lau2007state}
A.~W.~C. Lau and T.~C. Lubensky.
\newblock State-dependent diffusion: {Thermodynamic} consistency and its path
  integral formulation.
\newblock {\em Phys. Rev. E}, 76(1):011123, 2007.

\bibitem{revuz1999continuous}
D.~Revuz and M.~Yor.
\newblock {\em Continuous martingales and Brownian motion}, volume 293.
\newblock Springer, 1999.

\bibitem{van1992stochastic}
N.~G. Van~Kampen.
\newblock {\em Stochastic processes in physics and chemistry}, volume~1.
\newblock North holland, 1992.

\bibitem{kleinhans2007continuous}
D.~Kleinhans and R.~Friedrich.
\newblock Continuous-time random walks: Simulation of continuous trajectories.
\newblock {\em Phys. Rev. E}, 76(6):061102, 2007.

\bibitem{gardiner1985handbook}
C.~Gardiner.
\newblock Stochastic methods: {A Handbook} for the {Natural} and {Social
  Sciences}.
\newblock {\em Springer Series in Synergetics}, 2009.

\bibitem{samoradnitsky1994stable}
G.~Samoradnitsky and M.~S. Taqqu.
\newblock {\em Stable non-Gaussian random processes: stochastic models with
  infinite variance}, volume~1.
\newblock CRC press, 1994.

\bibitem{gnedenko1954limit}
B.~V. Gnedenko and A.~N. Kolmogorov.
\newblock Limit distributions for sums of independent.
\newblock {\em Amer. J. Math.}, 105:28--35, 1954.

\bibitem{feller1971introduction}
W.~Feller.
\newblock An introduction to probability and its applications, {Vol. II}.
\newblock {\em Wiley, New York}, 1971.

\bibitem{sato1999levy}
K.~I. Sato.
\newblock L{\'e}vy processes and infinite divisibility, 1999.

\bibitem{bertoin1999subordinators}
J.~Bertoin.
\newblock Subordinators: examples and applications.
\newblock In {\em Lectures on probability theory and statistics}, pages 1--91.
  Springer, 1999.

\bibitem{schilling2012bernstein}
R.~L. Schilling, R.~Song, and Z.~Vondracek.
\newblock {\em Bernstein functions: theory and applications}, volume~37.
\newblock Walter de Gruyter, 2012.

\bibitem{meerschaert2015relaxation}
M.~M. Meerschaert and B.~Toaldo.
\newblock Relaxation patterns and semi-{Markov} dynamics.
\newblock {\em arXiv preprint arXiv:1506.02951}, 2015.

\bibitem{jacod714calcul}
J.~Jacod.
\newblock {Calcul Stochastique et Probl{\`e}mes de Martingales [Stochastic
  Calculus and Martingale Problems]}.
\newblock {\em Lecture Notes in Mathematics}, 714, 1979.

\bibitem{kobayashi2011stochastic}
K.~Kobayashi.
\newblock Stochastic calculus for a time-changed semimartingale and the
  associated stochastic differential equations.
\newblock {\em J. Theoret. Probab.}, 24(3):789--820, 2011.

\bibitem{kunita1997stochastic}
H.~Kunita.
\newblock {\em Stochastic flows and stochastic differential equations},
  volume~24.
\newblock Cambridge university press, 1997.

\bibitem{magdziarz2010path}
M.~Magdziarz.
\newblock Path properties of subdiffusion - {A} martingale approach.
\newblock {\em Stoch. Model.}, 26(2):256--271, 2010.

\bibitem{Orzel2011FKKE}
S.~Orzel and A.~Weron.
\newblock Fractional {Klein-Kramers} dynamics for subdiffusion and
  \uppercase{I}t{\^o} formula.
\newblock {\em J. Stat. Mech.}, 2011, 2011.

\bibitem{lubashevsky2009realization}
I.~Lubashevsky, R.~Friedrich, and A.~Heuer.
\newblock Realization of {L}{\'e}vy walks as {Markovian} stochastic processes.
\newblock {\em Phys. Rev. E}, 79(1):011110, 2009.

\bibitem{lubashevsky2009continuous}
I.~Lubashevsky, R.~Friedrich, and A.~Heuer.
\newblock Continuous-time multidimensional {Markovian} description of {L\'e}vy
  walks.
\newblock {\em Phys. Rev. E}, 80(3):031148, 2009.

\bibitem{evans1993probability}
D.~J. Evans, E.~G.~D. Cohen, and G.~P. Morriss.
\newblock Probability of second law violations in shearing steady states.
\newblock {\em Phys. Rev. Lett.}, 71(15):2401, 1993.

\bibitem{gallavotti1995dynamical}
G.~Gallavotti and E.~G.~D. Cohen.
\newblock Dynamical ensembles in nonequilibrium statistical mechanics.
\newblock {\em Phys. Rev. Lett.}, 74(14):2694, 1995.

\bibitem{gallavotti1995reversible}
G.~Gallavotti.
\newblock Reversible {Anosov} diffeomorphisms and large deviations.
\newblock {\em Math. Phys. Electron. J.}, 1(1), 1995.

\bibitem{kurchan1998fluctuation}
J.~Kurchan.
\newblock Fluctuation theorem for stochastic dynamics.
\newblock {\em J. Phys. A}, 31(16):3719, 1998.

\bibitem{lebowitz1999gallavotti}
J.~L. Lebowitz and H.~Spohn.
\newblock A {Gallavotti--Cohen}-type symmetry in the large deviation functional
  for stochastic dynamics.
\newblock {\em J. Stat. Phys.}, 95(1-2):333--365, 1999.

\bibitem{wang2002experimental}
G.~M. Wang, E.~M. Sevick, E.~Mittag, D.~J. Searles, and D.~J. Evans.
\newblock Experimental demonstration of violations of the second law of
  thermodynamics for small systems and short time scales.
\newblock {\em Phys. Rev. Lett.}, 89(5):050601, 2002.

\bibitem{van2003stationary}
R.~Van~Zon and E.~G.~D. Cohen.
\newblock Stationary and transient work-fluctuation theorems for a dragged
  {Brownian} particle.
\newblock {\em Phys. Rev. E}, 67(4):046102, 2003.

\bibitem{trepagnier2004experimental}
E.~H. Trepagnier, C.~Jarzynski, F.~Ritort, G.~E. Crooks, C.~J. Bustamante, and
  J.~Liphardt.
\newblock Experimental test of {Hatano} and {Sasa}'s nonequilibrium
  steady-state equality.
\newblock {\em Proc. Natl. Acad. Sci.}, 101(42):15038--15041, 2004.

\bibitem{taniguchi2007onsager}
T.~Taniguchi and E.~G.~D. Cohen.
\newblock {Onsager-Machlup} theory for nonequilibrium steady states and
  fluctuation theorems.
\newblock {\em J. Stat. Phys.}, 126(1):1--41, 2007.

\bibitem{taniguchi2008nonequilibrium}
T.~Taniguchi and E.~G.~D. Cohen.
\newblock Nonequilibrium steady state thermodynamics and fluctuations for
  stochastic systems.
\newblock {\em J. Stat. Phys.}, 130(4):633--667, 2008.

\bibitem{gomez2010steady}
J.~R. Gomez-Solano, L.~Bellon, A.~Petrosyan, and S.~Ciliberto.
\newblock Steady-state fluctuation relations for systems driven by an external
  random force.
\newblock {\em EPL}, 89(6):60003, 2010.

\bibitem{aquino2013power}
J.~I. Jim\'enez-Aquino and R.~M. Velasco.
\newblock Power fluctuation theorem for a {Brownian} harmonic oscillator.
\newblock {\em Phys. Rev. E}, 87:022112, Feb 2013.

\bibitem{mestres2014realization}
P.~Mestres, I.~A. Martinez, A.~Ortiz-Ambriz, R.~A. Rica, and E.~Roldan.
\newblock Realization of nonequilibrium thermodynamic processes using external
  colored noise.
\newblock {\em Phys. Rev. E}, 90(3):032116, 2014.

\bibitem{touchette2007fluctuation}
H.~Touchette and E.~G.~D. Cohen.
\newblock Fluctuation relation for a {L\'e}vy particle.
\newblock {\em Phys. Rev. E}, 76(2):020101, 2007.

\bibitem{touchette2009anomalous}
H.~Touchette and E.~G.~D. Cohen.
\newblock Anomalous fluctuation properties.
\newblock {\em Phys. Rev. E}, 80(1):011114, 2009.

\bibitem{baule2009steady}
A.~Baule and E.~G.~D. Cohen.
\newblock Steady-state work fluctuations of a dragged particle under external
  and thermal noise.
\newblock {\em Phys. Rev. E}, 80(1):011110, 2009.

\bibitem{chechkin2009fluctuation}
A.~V. Chechkin and R.~Klages.
\newblock Fluctuation relations for anomalous dynamics.
\newblock {\em J. Stat. Mech.}, 2009(03):L03002, 2009.

\bibitem{chechkin2012normal}
A.~V. Chechkin, F.~Lenz, and R.~Klages.
\newblock Normal and anomalous fluctuation relations for {Gaussian} stochastic
  dynamics.
\newblock {\em J. Stat. Mech.}, 2012(11):L11001, 2012.

\bibitem{dieterich2015fluctuation}
P.~Dieterich, R.~Klages, and A.~V. Chechkin.
\newblock Fluctuation relations for anomalous dynamics generated by
  time-fractional fokker--planck equations.
\newblock {\em New J. Phys.}, 17(7):075004, 2015.

\bibitem{tassieri2016microrheology}
M.~Tassieri.
\newblock {\em {Microrheology with Optical Tweezers: Principles and
  Applications}}.
\newblock Pan Stanford, 2016.

\bibitem{karatzas2012brownian}
I.~Karatzas and S.~Shreve.
\newblock {\em Brownian motion and stochastic calculus}, volume 113.
\newblock Springer Science \& Business Media, 2012.

\bibitem{hanggi1982stochastic}
P.~H{\"a}nggi and H.~Thomas.
\newblock Stochastic processes: {Time} evolution, symmetries and linear
  response.
\newblock {\em Phys. Rep.}, 88(4):207--319, 1982.

\bibitem{klimontovich1990ito}
Y.~L. Klimontovich.
\newblock Ito, {Stratonovich} and kinetic forms of stochastic equations.
\newblock {\em Physica A}, 163(2):515--532, 1990.

\bibitem{ash2000probability}
R.~B. Ash and C.~Doleans-Dade.
\newblock {\em Probability and measure theory}.
\newblock Academic Press, 2000.

\end{thebibliography}
\end{document}